\documentclass[10pt,apj,floats,floatfix,uperscriptaddress,
             showkeys,showpacs,groupedaddress,amsmath,amsthm,lipsum]{revtex4-2}
\usepackage{graphicx}
\usepackage{epstopdf}
\usepackage{subfig}
\usepackage{amsmath}
\usepackage{verbatim}
\usepackage{mathrsfs}
\usepackage{amsfonts}
\usepackage{latexsym}
\usepackage{color}
\usepackage{epsfig}
\usepackage[usenames,dvipsnames]{xcolor}
\usepackage{graphicx, graphics}
\usepackage{units}
\usepackage{envmath}
\usepackage{natbib}
\usepackage{ctable}
\usepackage{soul}
\usepackage{time}
\usepackage{mathtools}
\begin{document}
\title{ Inflation and primordial fluctuations in $f(Q,T)$ gravity }
 
\author{Parviz Goodarzi}\email{parviz.goodarzi@abru.ac.ir}
 \affiliation{ Department of Physics, Faculty of Basic Sciences, Ayatollah Boroujerdi University, Boroujerd, Iran}

\begin{abstract}

We investigate slow roll inflation and the creation of primordial density fluctuations in the framework of $f(Q,T)$ gravity. Our focus is on constraining the evolution of both the background and perturbations in this theory, specifically using the form $f(Q,T) = \alpha Q + g(T)$, where $g(T)$ is an arbitrary function of the trace of the stress-energy tensor $T$. 
We derive the Mukhanov-Sasaki equations for scalar and tensor perturbations, and by solving these equations in the slow-roll regime, we compute the power spectra and spectral index for both modes within the general functional framework of $g(T)$. 
In particular, we examine power law functional forms of $g(T)$ to establish the observational constraints associated with quadratic potential. By imposing constraints on the model's parameters, we obtain results that align closely with the Planck 2018 data and BAO data for the tensor-to-scalar ratio. Notably, a model that includes $g(T) = \beta T^2$  and a quadratic potential yields the best-fit values consistent with the spectral index and tensor-to-scalar ratio suggested by the Planck and BICEP2 results.

\end{abstract}

\maketitle
	
\tableofcontents

\section{Introduction}

Recently, there has been a growing interest in understanding the early Universe, leading to extensive research in both theoretical and observational aspects \citep{Hinshaw2013,Aghanim2021,Planck2018,WMAP2006}. 
One of the most intriguing topics regarding the early Universe is cosmological inflation \cite{Martin2014,linde1990,Barrow1990,Linde1983,Guth1981}.
However, inflation was primarily addressed to resolve certain perplexing issues within the standard model of cosmology such as the flatness, monopole and horizon problems; it has been a very successful framework for explaining the generation and evolution of the primordial fluctuations as seeds of the large-scale structures of the Universe \cite{Starobinsky1980,Albrecht1982,Mukhanov1992}.
 Inflation refers to a brief phase in the Universe's development characterized by the accelerated expansion of the universe \cite{linde1982}.
Conventional inflation is defined by a de-Sitter expansion driven by an inflaton field, which has negligible kinetic energy compared to its potential energy \cite{Encyclo2014}.

After understanding the evolution of the cosmic background, the next step is to ascertain the evolution of cosmological perturbations. Inflation creates a mechanism that produces a Gaussian and nearly scale-invariant spectrum of primordial perturbations, aligning with observational data \cite{mukhanov}.
The examination of perturbed field equations through the decomposition of linear perturbations into scalar, vector, and tensor modes has enhanced our understanding of the quantum fluctuations as the origin of large-scale structures in the Universe \cite{Weinberg}.
Since the perturbations are assumed to be small to maintain the global homogeneity and isotropy of Robertson-Walker geometry, the second-order perturbative terms associated with the background are usually deemed negligible. Consequently, the linear terms are typically adequate to account for the slight deviations from the background.

Therefore, the physics of inflation has been extensively examined through a variety of gravitational models \cite{liddle1994,Odintsov2023,Martin2024}. One of the most successful models is $f(R)$ gravity, which encompasses Starobinsky inflation \cite{Starobinsky2007,Sotiriou2010,Felice2010,Vilenkin1985,Majic1986}, as well as generalized scalar-tensor models that encompass minimally and non-minimally coupled scalar fields models \cite{Fuji2003,Horndeski1974,Sadjadi2012,Sadjadi2013,Goodarzi2022}.
Furthermore, scalar field models have been used as the predominant prototype for inflationary models such as non-minimally coupled Higgs inflation \cite{Steinwachs}, T-model and E-model $\alpha$-attractors, and the D-brane KKLT potential \cite{Baumann2008,Kallosh2013,Kallosh2013b}.

Furthermore, an interesting strategy for addressing inflation is to adjust General Relativity by investigating geometries that go beyond the Riemannian framework. This includes examining torsion and non-metricity, which provide more advanced geometric descriptions of gravity \cite{Carrol2005,myrzak2015}.

Weyl, Dirac, Cartan, and Weitzenböck made efforts to develop a broader geometry beyond Riemannian geometry \cite{Weyl1918, Wheeler2018, Dirac1973}. In Weyl's theory, the covariant divergence of the metric tensor is not zero
 ($\bigtriangledown_\lambda g_{\mu\nu}\neq 0$), where $g_{\mu\nu}$ and this characteristic can be mathematically represented by a new geometric concept known as non-metricity $Q_{\lambda\mu\nu}=\bigtriangledown_\lambda g_{\mu\nu}$. Another important development of general relativity that has physical application is the Weitzenböck manifold, characterized by the properties $\bigtriangledown_\lambda g_{\mu\nu}=0$, $T^\lambda_{\mu\nu}\neq$, and  $R^\lambda_{\mu\nu\sigma}=0$ where $T^\lambda_{\mu\nu}$ and $R^\lambda_{\mu\nu\sigma}$ are the torsion tensor and curvature tensor respectively. Moreover, teleparallel gravity employs the Weitzenböck connection, characterized by zero curvature and non-metricity tensors, while exhibiting non-zero torsion \cite{Maluf2013}. This class of theory are called teleparallel gravity equivalence to general relativity (TEGR).
Additionally, symmetric teleparallel gravity utilizes a connection that is defined by having both zero curvature and torsion tensors, while also incorporating a non-metricity tensor to depict gravitational interactions. \cite{Naster1999, Jimenez2018}. Symmetric teleparallel gravity provides an alternative geometric framework for understanding gravity that is dynamically equivalent to General Relativity \cite{Lu2019, Jimenez2020, Dimakis2022}.
This theory is known as symmetric teleparallel gravity equivalent to general relativity (STEGR).
Another category of modified gravity theories that has recently been explored in the literature to explain inflation and dark energy includes modifications of TEGR and STEGR theories, which lead to $f(\tau)$ gravity and $f(Q)$ gravity, where $Q$ and $\tau$ are the non-metricity and torsion scalar respectively. \cite{Maluf2013,Bahamonde,Capozziello}

It is worth noticing that, the equivalence of $f(\tau)$ and $f(Q)$ with $f(R)$ holds only for theories linear in the scalar invariants; in general, the equivalence among the three representations is not valid. Extensions can involve further degrees of freedom, which lead to the breaking of equivalence among different representations of gravity. Specifically, in $f(R)$ gravity, the field equations are of fourth order in the metric representation, while $f(\tau)$ and $f(Q)$ remain second-order \cite{Vittorio2022}. 
This provides us with the strongest motivation to investigate inflation and primordial fluctuations in a new extension of $f(Q)$ represented as $f(Q,T)$ where $T$ is the trace of stress-momentum tensor.
$f(Q,T)$ gravity is an extended version of symmetric teleparallel gravity theory that includes the interaction between geometry and the trace of the energy-momentum tensor $T$ \cite{Xu2019}.

This coupling between geometry and matter enables an interesting description, such as the violation of stress-energy conservation, which may be described particle production induced by gravity in this framework \cite{Xu2019,Arora2020,Arora:2021}. An interesting aspect of these theories is their potential to interpretation as an effective framework for describing particular phenomena in quantum gravity \cite{Yang2016}.
Consequently, various studies and multiple cosmological features in the framework of $f(Q,T)$ gravity have been explored in both the early and late time acceleration of the Universe  \cite{Xu2020,Arora2021a,Gadbail2021,Godani2021,Pradhan2021,Bhagat2023,Shiravand2024}.
Novel couplings between non-metricity and matter in the form of $f(Q)L_M$ studied in Ref. \cite{Harko2018}. 
Minimal coupling in the presence of non-metricity and torsion is investigated in \cite{Delhom2020}.
In reference \cite{Arora2020a}, the acceleration and deceleration parameters for the model $f(Q,T)= mQ^n+ bT$ have been determined to be consistent with the current accelerating universe. This model has been tested using BAO and SNeIa data, demonstrating that $f(Q,T)$ models can be a geometric description of dark energy. Cosmological inflation within the context of $f(Q,T)$ gravity has been considered for specific linear version of $f(Q,T)$ gravity \cite{Shiravand2022}. They used the equivalence of the linear version of $f(Q,T)$ gravity with the linear version of $f(R,T)$ gravity to explore observational constraints \cite{Fisher2019}.
Also The exploration of cosmological linear perturbations in $f(Q,T)$ gravity, encompassing scalar, vector, and tensor decomposition, has been discussed in the Ref. \cite{Antonio2022}.
Notably, the function $f(Q,T)=-{(Q+2\Lambda)}/G_N-((16\pi)^2 G_Nb)/(120 H_0^2)T^2$ has demonstrated a significant preference against $\Lambda CDM$ when analyzed using SNeIa data.
This findings encourages further exploration of $f(Q, T)$ beyond the background framework \cite{Antonio2021,Antonio2022}.

Despite extensive studies conducted on this model, generation of primordial fluctuations in cold inflation have yet to be examined within the context of $f(Q,T)$ gravity.
Therefore, this has motivated us to consider symmetric teleparallel gravity, where non-metricity $Q$ is coupled with the general function of the trace of the energy-momentum tensor $g(T)$.
In present paper, we intend to study carefully inflation and generation of primordial fluctuations within the framework of $f(Q,T)$ gravity, specifically focusing on the general form $f(Q,T)=\alpha Q+g(T)$, where $g(T)$ is an arbitrary function of the trace of the stress-energy tensor $T$ and $\alpha$ is a constant. 
By examining the first-order cosmological perturbations, we will derive the Mukhanov-Sasaki equation for both scalar and tensor modes in $f(Q,T)$ gravity.
Additionally, by solving the Mukhanov-Sasaki equations, using the slow roll approximation, we will derive the power spectrum $\mathcal{P}_s$, the spectral index $n_s$, and the tensor-to-scalar ratio $r$ for a general function of the trace of the stress-energy tensor, denoted as $g(T)$.
Furthermore, we compare our results with the observational data from Planck 2018 for the power law functional form of $g(T)$, expressed as $g(T)=\beta T^n $ for different values of power $n$.
By imposing constraints on the parameters of the model, we achieve results that align well with the Planck 2018 data. Interestingly, a model including a $T^2$ dependence showed good agreement with the observational data.

The structure of the work is organized as follows: 
In Section II, we examine $f(Q,T)$ gravity and derive the background equations of motion within the FLRW spacetime.
In Section III, we consider cosmological inflation within the framework of $f(Q,T)=\alpha Q+g(T)$, where $g(T)$ represents a general function of the trace of the stress-energy tensor.
In Section IV, we explore the cosmological perturbations arising from the function $f(Q,T) = \alpha Q+g(T)$ within the context of slow roll inflation.
In Section V, we applied the equations developed in previous chapters to the generic function $g(T)$ in specific cases $f(Q,T) = \alpha Q+\beta T$ and $f(Q,T)=\alpha Q+\beta T^2$. 
In Section VI, we explore our results numerically. 
Finally, Section VII is devoted to the concluding findings obtained.

\section{Equation of motion in $f(Q,T)$ Gravity}

Initially, we will provide a concise overview of $f(Q,T)$ gravity and derive the background equations of motion within the Friedmann-Lemaiter-Robertson-Walker (FLRW) spacetime. The most general action for $f(Q,T)$ gravity in the
presence of the inflaton field is given by \cite{Xu2019}
\begin{equation}\label{1}
S=\int \sqrt{-g}\left[\dfrac{f(Q,T)}{2\kappa}+L^{[\phi]}\right]{\rm d}^{4}x,
\end{equation}
where $g$ is the determinant of the metric tensor $g_{\mu\nu}$ and $\kappa=8\pi G$ is the gravitational coupling constant. The gravitational Lagrangian density is given by $f(Q,T)$, where $f$ is an arbitrary function of the non-metricity
  $Q$ and $T$ is the trace of energy-momentum tensor $T=g^{\mu\nu}T_{\mu\nu}$. 
$L^{[\phi]}$ represents the Lagrangian of matter, which in our analysis is a scalar field.
In this action dynamical variables are the inflaton field, the metric
and the connection, which we assume to be in the case of vanishing
the Riemann tensor and torsion tensor.

In Weyl-Cartan geometry, it is established that the connection can be broken down into three distinct components, namely \cite{Xu2019,Antonio2022}
\begin{equation}\label{2}
\Gamma^{\alpha}{}_{\mu\nu}=\{^{\alpha}{}_{\mu\nu}\}+C^{\alpha}{}_{\mu\nu}+L^{\alpha}{}_{\mu\nu},
\end{equation}
where $\{^{\alpha}{}_{\mu\nu}\}$, $C^{\alpha}{}_{\mu\nu}$ and
$L^{\alpha}{}_{\mu\nu}$ represent the Christoffel symbol, the
contorsion tensor and the disformation tensor, respectively.
In the standard definition, we can express these three connections as follows:
\begin{eqnarray}\label{3}
\{^{\alpha}{}_{\mu\nu}\}&=&\dfrac{1}{2}g^{\alpha\beta}(\partial_{\mu}
g_{\beta\nu} +\partial _{\nu}g_{\beta\mu}-\partial_{\beta}
g_{\mu\nu}),\\
C^{\alpha}{}_{\mu\nu}&=&\dfrac{1}{2}\mathbb{T}^{\alpha}{}_{\mu\nu}+\mathbb{T}_{(\mu\nu)}{}^{\alpha},\\
L^{\alpha}{}_{\mu\nu}&\equiv
&-\dfrac{1}{2}g^{\alpha\beta}\left(\nabla_{\mu}g_{\beta\nu}
+\nabla_{\nu}g_{\beta\mu}-\nabla_{\beta}g_{\mu\nu}\right),
\end{eqnarray}
where $\mathbb{T}^{\alpha}{}_{\mu\nu}=
\Gamma^{\alpha}{}_{\mu\nu}-\Gamma^{\alpha}{}_{\nu\mu}$ represents the torsion tensor, $Q_{\alpha\mu\nu}\equiv\nabla_{\alpha} g_{\mu\nu}\neq 0$ is the covariant derivative of the metric tensor with respect to the Weyl-Cartan connection $\Gamma^{\alpha}{}_{\mu\nu}$, stands for the non-metricity tensor.
Also, we define the traces of non-metricity tensor
 ${Q}_{\alpha}\equiv Q_{\alpha}{}^{\mu}{}_{\mu}$ and $\tilde{Q}_{\alpha}\equiv Q^{\mu}{}_{\alpha\mu}$, the super potential tensor and the non-metricity scalar are given by
\begin{equation}\label{4}
P^{\alpha}{}_{\mu\nu}\equiv
-\dfrac{1}{2}L^{\alpha}{}_{\mu\nu}+\dfrac{1}{4}
\left(Q^{\alpha}-\tilde{Q}^{\alpha}\right)g_{\mu\nu}-\dfrac{1}{4}\delta^{\alpha}_{(\mu}Q_{\nu)},
\end{equation}
\begin{equation}\label{5}
Q\equiv -g^{\mu
\nu}\left(L^{\alpha}{}_{\beta\mu}L^{\beta}{}_{\nu\alpha}-L^{\alpha}{}_{\beta\alpha}L^{\beta}{}_{\mu\nu}\right).
\end{equation}
By utilizing the action variational principle, $\delta S = 0$, the variation of the action \eqref{1} with respect to the metric tensor $g_{\mu\nu}$ results in the modified gravitational field equations, as follows \cite{Xu2019}:
\begin{eqnarray}\label{6}
\kappa\,T_{\mu\nu}^{[\phi]}&=&
-\dfrac{2}{\sqrt{-g}}\bigtriangledown_{\alpha}
\left(f_{Q}\sqrt{-g}P^{\alpha}{}_{\mu\nu}\right) \nonumber\\
&-&\dfrac{f}{2}g_{\mu\nu}
+f_T\left(T_{\mu\nu}^{[\phi]}+\Theta_{\mu\nu}\right) \nonumber\\
&-&f_Q\left(P_{\mu\alpha\beta} Q_{\nu}{}^{\alpha\beta}-2\,
Q^{\alpha\beta}{}_{\mu}P_{\alpha\beta\nu}\right),
\end{eqnarray}
where the auxiliary tensor $\Theta_{\mu\nu}$, together with $f_Q$ and $f_T$ are defined as 
\begin{equation}\label{7}
\Theta_{\mu\nu}\equiv g^{\alpha\beta}\dfrac{\delta
T_{\alpha\beta}^{[\phi]}} {\delta g^{\mu\nu}},~~ f_{\rm
Q}\equiv\dfrac{\partial f(Q,T)}{\partial Q},~~ f_{\rm
T}\equiv\dfrac{\partial f(Q,T)}{\partial T}.
\end{equation}
The scalar field energy-momentum tensor $T^{[\phi]}_{\mu\nu}$, is defined in terms of variation of Lagrangian of scalar field with respect to metric as
\begin{equation}\label{8}
T^{[\phi]}_{\mu\nu}\equiv-\dfrac{2}{\sqrt{-g}}\frac{\delta(\sqrt{-g}
L^{[\phi]})}{\delta g^{\mu\nu}}.
\end{equation}
Furthermore, we can express the field equations \eqref{6} in an alternative format 
\begin{eqnarray}\label{8.1}
&f_{T}& \left(T^{[\phi]\mu}{}_{\nu}+\Theta^{\mu}{}_{\nu}\right)-\kappa T^{[\phi]\mu}{}_{\nu} \nonumber\\
&=&\dfrac{f}{2}\delta^{\mu}{}_{\nu}+f_{Q}
Q_{\nu}{}^{\alpha\beta} P^{\mu}{}_{\alpha\beta}
+\dfrac{2}{\sqrt{-g}}\nabla_\alpha\left(f_{Q} \sqrt{-g}
P^{\alpha\mu}{}_{\nu}\right).
\end{eqnarray}
To investigate the conservation of the energy-momentum tensor in the $f(Q,T)$ gravity, we take the covariant divergence of $T_{\mu\nu}^{[\phi]}$ with respect to connection as follows \cite{Antonio2022,Xu2019}:
\begin{equation}\label{9}
\mathcal{D}_{\mu} T^{[\phi]\mu}{}_{\nu}-\dfrac{1}{f_{T}+\kappa}\left(f_{T} \partial_\nu p^{[\phi]}
-\dfrac{1}{2}f_{T}
\partial_\nu T^{[\phi]}-B_\nu \right)=0,
\end{equation}
where $f_{T}+\kappa\neq 0$, the symbol $\mathcal{D}_{\mu}$
signifies the covariant derivative with respect to the Levi-Civita
connection, and $B_{\nu}$ in this relation defined as
\begin{equation}\label{10}
B_{\nu}\equiv \dfrac{1}{\sqrt{-g}}\Big[Q_\mu \nabla_\alpha \Big(\sqrt{-g}f_{\rm Q} P^{\alpha\mu}{}_{\nu}\Big)+2 \nabla_\mu \nabla_\alpha\Big(\sqrt{-g}f_{Q}
P^{\alpha\mu}{}_{\nu}\Big)\Big].
\end{equation}
Additionally, by varying the action \eqref{1} with respect to the connection, we derive another set of field equations as follows:
\begin{equation}\label{11}
\bigtriangledown_{\mu}\bigtriangledown_{\nu}\(2\sqrt{-g}f_{Q}P^{\mu\nu}{}_{\alpha}+\kappa\, {\cal H}_{\alpha}{}^{\mu\nu}\)=0,
\end{equation}
where ${\cal H}_{\alpha}{}^{\mu\nu}$ is the hypermomentum tensor density defined as
\begin{equation}\label{12}
{\cal H}_{\alpha}{}^{\mu\nu}\equiv\dfrac{f_{T}\sqrt{-g}}{2\kappa}\dfrac{\delta\,T^{[\phi]}}{\delta\Gamma^{\alpha}{}_{\mu\nu}}+\dfrac{\delta\(\sqrt{-g}\,L_{m}^{[\phi]}\)}{\delta \Gamma^{\alpha}{}_{\mu\nu}}.
\end{equation}
We will now examine $f(Q,T)$ gravity within the framework of the flat FLRW metric, which is represented in the form of 
\begin{equation}\label{13}
ds^2=-dt^2+a^2(t)\delta_{ij}dx^idx^j,
\end{equation}
where $a(t)$ represents the scale factor.
We assume that matter component of the universe is perfect fluid. Consequently, the field equations \eqref{6} provide the energy density and pressure of the scalar field as follows:
\begin{eqnarray}
\label{14}
\kappa\rho^{[\phi]}&=&\dfrac{f}{2}-6f_{Q}H^2-\dfrac{2f_{T}}{\kappa+f_{T}}\left(\dot{f}_{Q}H+f_{Q}\dot{H}\right),\\
\label{14.1}
\kappa p^{[\phi]}&=&-\dfrac{f}{2}+6f_QH^2+2\left(\dot{f}_QH+f_Q\dot{H}\right),
\end{eqnarray}
where $H$ is the Hubble parameter defined as $H=\dot{a}/a$. In the following an overdot denoted the derivative with respect to cosmic time $t$. Also, we can write equations \eqref{14} and \eqref{14.1} as a modified Friedmann equations as
\begin{eqnarray}
\label{15}
3H^2&=&\dfrac{f}{4f_Q}-\dfrac{1}{2f_Q}\left[\left(\kappa+f_T\right)\rho^{[\phi]}+f_Tp^{[\phi]}\right],\\
\label{15.1}
2\dot{H}+3H^2&=&\dfrac{1}{2f_Q}\left[\left(\kappa+f_T\right)\rho^{[\phi]}+ \left(2\kappa+f_T\right)p^{[\phi]}\right] \nonumber \\
&+&\dfrac{f}{4f_Q}-\dfrac{2\dot{f}_Q}{f_Q}H.
\end{eqnarray}
Furthermore, we can obtain the evolution of the Hubble parameter $H$ by combining equations \eqref{14} and \eqref{14.1} in the following manner:
\begin{equation} \label{16}
\dot{H}+\frac{\dot{f}_Q}{f_Q}H=\frac{1}{2f_Q}\left(\kappa+f_T\right)
\left(\rho^{[\phi]}+p^{[\phi]}\right).
\end{equation}
Finally, the energy balance equation in $f(Q,T)$ gravity is derived from relation \eqref{9} within the FLRW metric
\begin{eqnarray}\label{17}
\frac{\partial\rho^{[\phi]}}{\partial t}&+&3H\left(\rho^{[\phi]}+p^{[\phi]}\right)=\frac{f_T}{2(\kappa+f_T)(\kappa+2f_T)} \nonumber \\
&&\times\left(\dot{s}-\left(\frac{3f_T+2\kappa}{\kappa+f_T}\right)\frac{\dot{f_T}}{f_T}s+6Hs\right),
\end{eqnarray}
where we have denoted $s=2(\dot{f_Q}H+f_Q\dot{H})$.
It is clear that energy conservation is maintained when $f_T = 0$
Therefore, the presence of $T$ in the Lagrangian breaks the energy conservation. 

\section{Cosmological Inflation with $f(Q,T)=\alpha Q+g(T)$}
\subsection{The field equations}

In the previous section, we provided a brief overview of the general structure of $f(Q,T)$ gravity. Additionally, we derived the general field equations for $f(Q,T)$ gravity in the context of a spatially flat FLRW geometry.
In this section, we will examine cosmological inflation through the framework of $f(Q,T)$ gravity, focusing on the specific functional form $f(Q,T)=\alpha Q+g(T)$. Here, $g(T)$ denotes an arbitrary function of the trace of the energy-momentum tensor $T$, and $\alpha$ is a constant parameter of the model.
Additionally, we propose that the matter is represented by a canonical scalar field defined by the following Lagrangian:
\begin{equation}\label{18}
L^{[\phi]}=-\dfrac{1}{2}g^{\mu\nu}\partial_{\mu}\phi\,\partial_{\nu}\phi-V(\phi),
\end{equation}
where $V(\phi)$ is inflaton potential. By using $\phi\equiv\phi(t)$ as a homogeneous scalar field, we can easily obtain the energy-momentum tensor associated with the inflaton field $T^{[\phi]}_{\mu\nu}$. Therefore the energy density $\rho^{[\phi]}$ and pressure $p^{[\phi]}$ associated to scalar field becomes
\begin{equation}\label{19}
\rho^{[\phi]}=\dfrac{1}{2}\dot{\phi}^2+V(\phi),\qquad
p^{[\phi]}=\dfrac{1}{2}\dot{\phi}^2-V(\phi).
\end{equation}
Moreover, by using equation \eqref{19} the trace of the energy-momentum tensor $T$ can be written as
\begin{equation}\label{20}
T^{[\phi]}=\dot{\phi}^2-4V.
\end{equation}
To obtain the field equations of $f(Q,T)$ in the functional form $f(Q,T)=\alpha Q+g(T)$, we use $f_Q=\alpha$ and $f_T=g_{,T}=\dfrac{dg(T)}{dT}$.
The energy balance equation \eqref{17}, becomes
\begin{eqnarray}\label{21}
\frac{\partial\rho^{[\phi]}}{\partial t}&+&3H\left(\rho^{[\phi]}+p^{[\phi]}\right)=\frac{\alpha g_{,T}}{(\kappa+g_{,T})(\kappa+2g_{,T})} \nonumber\\
&&\left[\ddot{H}-\frac{\dot{g_{,T}}}{g_{,T}}\left(\frac{2\kappa+3g_{,T}}{\kappa+g_{,T}}\right)\dot{H}+6H\dot{H}\right].
\end{eqnarray}
By combining of equations \eqref{19} and \eqref{16} we can obtain 
\begin{equation}\label{22}
\rho^{[\phi]}+p^{[\phi]}=\dot{\phi}^2=\left(\dfrac{2\alpha}{\kappa+g_{,T}}\right)\dot{H}.
\end{equation}
Specifically, by substituting of equations \eqref{19} and \eqref{22} into equation \eqref{21} we can derive modified equation of motion for scalar field as follows
\begin{eqnarray}\label{23}
\ddot{\phi}\left(\kappa+g_{,T}\right)+3H\dot{\phi}\left(\kappa+g_{,T}\right)
&+&V^{\prime}\left(\kappa+2g_{,T}\right) \nonumber \\
&=&-\dot{g}_{,T}\dot{\phi}.
\end{eqnarray}
It has been noted that the term $\dot{g}_{,T}$ on the right side of this equation behaves similarly to the the friction term present in warm inflation. Consequently, in models that include non vanishing $\dot{g}_{,T}$ term, the energy density of the scalar field decreases more rapidly compared to standard inflation during the expansion of the universe.
The modified Friedmann equations \eqref{15} and \eqref{15.1} in this context becomes
\begin{eqnarray}
\label{23.1}
3H^2&=&\dfrac{g}{2\alpha}-\dfrac{1}{\alpha}\left[\left(\kappa+g_{,T}\right)\rho^{[\phi]}+g_{,T}p^{[\phi]}\right],\\
\label{23.2}
4\dot{H}+3H^2&=&\dfrac{g}{2\alpha} \nonumber \\
&+&\dfrac{1}{\alpha}\left[\left(\kappa+g_{,T}\right)\rho^{[\phi]}+(2\kappa+g_{,T})p^{[\phi]}\right].
\end{eqnarray}
Also, based on equations \eqref{14} and \eqref{14.1}, the energy density and pressure of the scalar field within this framework can be expressed as
\begin{eqnarray}
\label{24}
\rho^{[\phi]}&=&\dfrac{g}{2\kappa}-\dfrac{3\alpha}{\kappa}H^2-\dfrac{2\alpha}{\kappa}\left(\dfrac{g_{,T}}{\kappa+g_{,T}}\right)\dot{H},\\
\label{24.1}
p^{[\phi]}&=&-\dfrac{g}{2\kappa}+\dfrac{\alpha}{\kappa}(2\dot{H}+3H^2).
\end{eqnarray}
In the following subsection, we will examine slow-roll inflation within this cosmological context. 

\subsection{General setup for slow-roll}

In this subsection we will investigate slow roll inflation within the $f(Q,T)=\alpha Q+g(T)$ framework for the general expression of the function $g(T)$. In order to apply the slow-roll approximation to all background equations, we define the dimensionless slow-roll parameters in the following manner:
\begin{eqnarray}\label{24.2}
&\lambda\equiv\dfrac{\ddot{\phi}}{3H\dot{\phi}},
~~~~\eta\equiv\dfrac{\dot{\phi}^2}{2V(\phi)},
~~~~\theta\equiv\dfrac{\dot{g}_{,T}}{3H(\kappa+g_{,T})}, \nonumber \\
&\epsilon\equiv-\dfrac{\dot{H}}{H^2},
~~~~\delta\equiv\dfrac{\ddot{H}}{2\dot{H}H}.
\end{eqnarray} 
Slow-roll inflation takes place when all of these slow-roll parameters are significantly less than one, as follows:
\begin{equation}\label{25}
\{\lambda,\eta,\theta,\epsilon,\delta\} \ll 1.
\end{equation}
By utilizing the slow-roll approximation \eqref{25} in the modified field equation \eqref{23}, we obtain
\begin{equation}\label{26}
3H\dot{\phi}\left(\kappa+g_{,T}\right)
\approx -V^{\prime}\left(\kappa+2g_{,T}\right).
\end{equation}
The background equations indicate a linear relationship between the slow-roll parameters during inflation, enabling us to apply a linear approximation to any function of the background quantity. Consequently, we can express $g(T)$ using a first-order expansion in terms of the slow roll parameter
\begin{eqnarray}
\label{27}
g&=&g(T)=g({\dot{\phi}}^2-4V(\phi))\approx\tilde{g}+\dot{\phi}^2\tilde{g}_{,T},\\
\label{27.1}
g_{,T}&=&g_{,T}(T)=g_{,T}({\dot{\phi}}^2-4V(\phi))\approx\tilde{g}_{,T}+\dot{\phi}^2\tilde{g}_{,TT},
\end{eqnarray}
where in these relations we define $\tilde{g}=g(-4V(\phi))$,~ $\tilde{g}_{,T}=g_{,T}(-4V(\phi))$~ and~ $\tilde{g}_{,TT}=g_{,TT}(-4V(\phi))$.
By using equation \eqref{22} we can rewrite equation \eqref{24} in the following form
\begin{equation}\label{28}
\rho^{[\phi]}=\dfrac{g}{2\kappa}-\dfrac{3\alpha}{\kappa}H^2-\dfrac{g_{,T}}{\kappa}\dot{\phi}^2.
\end{equation}
By substituting equation \eqref{19} into equation \eqref{28} and applying equations \eqref{27} and \eqref{27.1} at the leading order, we arrive at the following result  
\begin{equation}\label{29}
H^2\approx\dfrac{\tilde{g}-2\kappa V}{6\alpha}.
\end{equation}
In the case where $g(T)=\beta T$, we determine that $\tilde{g}=-4\beta V(\phi) $. As a result, equation \eqref{29} yields the familiar relation $H^2=-(\alpha+2\beta)V/3\alpha$, which is typical of linear $f(Q,T)$ gravity \cite{Shiravand2022}.
At the leading order we can write $\dot{\phi}$ from equation \eqref{26} in the following form
\begin{equation}\label{30}
\dot{\phi}\approx
-\left(\dfrac{\kappa+2\tilde{g}_{,T}}{\kappa+\tilde{g}_{,T}}\right)
\dfrac{V^{\prime}}{3H}.
\end{equation}
Through the following calculations, the slow-roll parameters are expressed in terms of the inflaton potential: The parameter $\lambda$ is derived by differentiating equation \eqref{26}. The parameter $\eta$ is determined from equations \eqref{29} and \eqref{30}. We calculate $\theta$, using the relations $\dot{g}_{,T}=g_{,TT}\dot{T}$ and $\dot{T}=2\ddot{\phi}\dot{\phi}-4\dot{\phi}V^{\prime}$. Equation \eqref{30} is used to find $\epsilon$. Finally, $\delta$ is obtained by differentiating equation \eqref{25}, which yields
\begin{eqnarray}
\label{31}\lambda&\approx&\left(\dfrac{\kappa}{\kappa+2\tilde{g}_{,T}}\right)\theta+
\dfrac{2\alpha(\kappa+2\tilde{g}_{,T})}{3(\kappa+\tilde{g}_{,T})}
\dfrac{V^{\prime\prime}}{(2\kappa V-\tilde{g})}+\dfrac{\epsilon}{3},\\
\label{32}\eta&\approx&
-\left(\dfrac{\kappa+2\tilde{g}_{,T}}{\kappa+\tilde{g}_{,T}}\right)^2
\dfrac{\alpha{V^{\prime}}^2}{3V(2\kappa V-\tilde{g})},\\
\label{33}\theta&\approx&
-\dfrac{8\alpha(\kappa+2\tilde{g}_{,T})\tilde{g}_{,TT}}{3(\kappa+\tilde{g}_{,T})^2}\dfrac{{V^{\prime}}^2}{(2\kappa V-\tilde{g})},\\
\label{34}\epsilon&\approx&
-\dfrac{2\alpha(\kappa+2\tilde{g}_{,T})^2}{(\kappa+\tilde{g}_{,T})}
\dfrac{{V^{\prime}}^2}{(2\kappa V-\tilde{g})^2},\\
\label{35}\delta&\approx& -3\lambda-\dfrac{3}{2}\theta.
\end{eqnarray}
Another significant quantity of cosmological inflation is the number of e-folds, defined as \cite{Martin2014}
\begin{equation}\label{36}
\mathcal{N}=\int_{t_{\star}}^{t_{end}}H{dt}=
\int_{\phi_{\star}}^{\phi_{end}}\dfrac{H}{\dot{\phi}}{d\phi}.
\end{equation}
Here, the subscript `$end$' and $\star$ denote the values of the
quantities at the end of inflation and at the horizon crossing respectively.
By utilizing equations \eqref{29} and \eqref{30} in the slow roll approximation we can derive the e-folding number, in terms of the inflaton potential as  
\begin{equation}\label{37}
\mathcal{N}\approx\int_{\phi_{\star}}^{\phi_{end}}
\dfrac{(\kappa+\tilde{g}_{,T})(2\kappa V-\tilde{g})}
{2\alpha(\kappa+2\tilde{g}_{,T})V^{\prime}}d\phi.
\end{equation}
Moreover, the significant quantity of background, during inflation, is the equation of state, which can be defined as
\begin{equation}\label{37.1}
\bar{\omega}\equiv\dfrac{p^{[\phi]}}{\rho^{[\phi]}}=\dfrac{\dot{\phi}^2/2-V(\phi)}{\dot{\phi}^2/2+V(\phi)},
\end{equation}
In the slow roll approximation, we can derive the equation of state in the following form: 
\begin{equation}\label{38}
\bar{\omega}\approx-1+2\eta.
\end{equation}
Also, the sound speed $\bar{c_s}^2$ is given by 
\begin{equation}\label{39}
\bar{c_s}^2\equiv\dfrac{\dot{p}^{[\phi]}}{\dot{\rho}^{[\phi]}}=
\dfrac{\ddot{\phi}-V^{\prime}}{\ddot{\phi}+V^{\prime}},
\end{equation}
where we can derive the sound speed corresponding functional form of $f(Q,T)$ gravity in the form $f(Q,T)=\alpha Q+g(T)$ as a function of slow roll parameters becomes
\begin{equation}\label{40}
\bar{c_s}^2\approx -1-2\left(\dfrac{\kappa+2\tilde{g}_{,T}}{\kappa+\tilde{g}_{,T}}\right)\lambda.
\end{equation}
 
Inflation begins at time $t_{\star}$, when the slow-roll parameter $\epsilon$ is much less than one. As time goes on, $\epsilon$ steadily rises until it equals one at the end of inflation. Consequently, during the inflationary period, $\epsilon$ must remain positive ($\epsilon>0$), which imposes certain constraints on the results obtained as follows
\begin{equation}\label{40.1}
\dfrac{-2\alpha}{\kappa+\tilde{g}_{,T}}>0.
\end{equation}

\section{Cosmological perturbations}

In the previous sections , we studied a homogeneous and isotropic universe within the framework of $f(Q,T)$ gravity. While the universe appears homogeneous and isotropic on a large scale, we know that there are deviations from these assumptions at smaller scales. In the cold inflationary model, these cosmological perturbations originate from the quantum fluctuations of the inflaton field, and these fluctuations constitute the initial conditions for the formation of large-scale structures \cite{Weinberg,mukhanov}.
In the following, we will establish the essential equations for scalar and tensor perturbations within the framework of $f(Q,T)$ gravity. Next, we explore the perturbed field equations within the context of $f(Q,T)$ gravity, specifically in the form $f(Q,T) = \alpha Q + g(T)$. Finally, we compute observable quantities, including the spectral index and power spectrum for both scalar and tensor perturbations.

\subsection{Scalar perturbations in $f(Q,T)$ gravity}
In this subsection, we explore the evolution equations of the cosmological perturbations within the $f(Q,T)$ framework, which includes a canonical scalar field. Additionally, we concentrate specifically on second-order perturbations. Following the conventions established in the literature, we will now use a bar over unperturbed quantities to distinguish them from perturbed quantities.
 The most general form of the scalar perturbations in the line element becomes
\begin{eqnarray}\label{41}
ds^2=-(1+2\Psi)dt^2+a^2(t)[(1-2\Phi)\delta_{ij} \nonumber \\
-2\partial_i\partial_jE]dx^idx^j+a(t)\partial_iWdt dx^i.
\end{eqnarray}
We can divide the metric into two components: the background metric and the perturbations in the following manner
\begin{equation}\label{41.1}
g_{\mu\nu}=\bar{g}_{\mu\nu}+h_{\mu\nu},
\end{equation}
where the unperturbed or background metric is the FLRW metric \eqref{13} and, the perturbed parts of the metric can be written as
\begin{eqnarray}\label{42}
h_{\mu\nu}=\begin{bmatrix}
-2\Psi & a\partial_iW \\
a\partial_iW &2a^2(-\Phi\delta_{ij}+\partial_i\partial_jE) \\
\end{bmatrix}.
\end{eqnarray}
As previously stated, we regard the universe as containing a perfect fluid, i.e. 
\begin{equation}\label{43}
T_{\mu\nu}=\(\rho^{[\phi]}+p^{[\phi]}\) u_{\mu}u_{\nu}+p^{[\phi]}g_{\mu\nu},
\end{equation}
were $u_{\mu}$ is the four velocity vector contains background parts and perturbed parts given by \cite{Weinberg}
\begin{equation}\label{43.1}
u_{\mu}(t,{\bf x})=\bar{u}_{\mu}(t)+\delta u_{\mu}(t,{\bf x}).
\end{equation}
The components of the unperturbed four-velocity of the perfect fluid, denoted as $\bar{u}_{\mu}$, are given by $\bar{u}_0 = -1$ and $\bar{u}_i = 0$. This is consistent with the normalization condition $u^{\alpha}u_{\alpha}=g^{\mu\nu}u_{\mu}u_{\nu} = -1$ in the rest frame.
Furthermore, by adding a scalar perturbations to the stress-energy tensor, we derive
\begin{eqnarray}\label{43.2}
T^\mu_{~~\nu}=\begin{bmatrix}
-\bar{\rho}^{[\phi]}-\delta\rho^{[\phi]} 
&\(\bar{\rho}^{[\phi]}+\bar{p}^{[\phi]}\)\partial_i\delta u\\
\(\bar{\rho}^{[\phi]}+\bar{p}^{[\phi]}\){\partial_i\(\delta u-W\)\over -a} 
&\(\bar{p}^{[\phi]}+\delta p^{[\phi]}\)\delta_{ij} \\
\end{bmatrix},
\end{eqnarray}
where $\delta u$ is scalar velocity potential defined as $\delta u_i=a(t)\partial_i\delta u$.
To perturb the scalar field and energy density, we decompose them into a background component and a perturbation component as follows:
\begin{eqnarray}\label{43.3}
\phi(t,{\bf x})&=&\bar{\phi}(t)+\delta \phi(t,{\bf x}),\\
\rho^{[\phi]}(t,{\bf x})&=&\bar{\rho}^{[\phi]}(t)+\delta
\rho^{[\phi]}(t,{\bf x}).
\end{eqnarray}
We now require a set of equations to examine the evolution of the perturbed quantity. By substituting the perturbed metric and relevant quantities into the field equations \eqref{8.1}, one can derive the perturbed field equations \cite{Antonio2022}.
Consequently, by perturbing the field equation \eqref{8.1} and taking the $0-0$ component, we obtain
\begin{align}\label{44}
&\delta\rho^{[\phi]}\Big[\kappa+{3\over2}\bar{f}_{
T}-\bar{f}_{\rm TT}\Big(\bar{\rho}^{[\phi]}
+\bar{p}^{[\phi]}\Big)\Big] \nonumber \\
&+\delta p^{[\phi]}\Big[\!-{1\over2}\bar{f}_{T}+3\bar{f}_{TT}\(\bar{\rho}^{[\phi]}+\bar{p}^{[\phi]}\)\!\Big]\nonumber\\
&=6H\(\bar{f}_{Q}+12H^2\bar{f}_{QQ}\)\(H\Psi+\dot{\Phi}\)-2\bar{f}_Qa^{-2}\nabla^2\Phi \nonumber \\
&+2H\[\bar{f}_Q+3\bar{f}_{QQ}\(2H^2+\dot{H}\)\]a^{-1}\nabla^2W,
\end{align}
where $\nabla^2=\Sigma_{i=1}^3(\partial/\partial x^i)^2$ is the Laplacian, $\bar{f}_{QQ}=\partial \bar{f}_{Q}/\partial Q$ and $\bar{f}_{TT}=\partial
\bar{f}_{T}/\partial T$.
Furthermore, by perturbing the field equation \eqref{8.1} and taking the trace of the spacial part of it, we have
\begin{align}\label{45}
&{\dfrac{1}{4}}\bar{f}_{T}\delta\rho^{[\phi]}-\Big(\frac{\kappa}{2}
+{3\over4}\bar{f}_{T}\Big)\delta p^{[\phi]}\nonumber\\
&=\(\bar{f}_{Q}+12H^2\bar{f}_{QQ}\)\(H\dot{\Psi}+H\dot{\Phi}+\ddot{\Phi}\)\nonumber\\
&+\[\bar{f}_{Q}\(3H^2+\dot{H}\)
+12\bar{f}_{QQ}H^2\(3H^2+4\dot{H}\)+12\dot{\bar{f}}_{QQ}H^3\]\Psi
 \nonumber \\
&+\dfrac{1}{3}\bar{f}_{Q}a^{-2}\nabla^2\Psi \nonumber \\
&+2\[\bar{f}_{Q}+6\bar{f}_{QQ}\(2H^2+3\dot{H}\)+6H\dot{\bar{f}}_{QQ}\]H\dot{\Phi}-\dfrac{1}{3}\bar{f}_{Q}a^{-2}\nabla^2\Phi \nonumber\\
&+\dfrac{1}{3}\[2\bar{f}_{Q}+3\bar{f}_{QQ}\(4H^2+5\dot{H}\)+6H\dot{\bar{f}}_{QQ}\]Ha^{-1}\nabla^2W \nonumber \\
&+\dfrac{1}{3}\(\bar{f}_{Q}+6\bar{f}_{QQ}H^2\)a^{-1}\nabla^2\dot{W} \nonumber \\
&-\[\bar{f}_{Q}+12\bar{f}_{QQ}\(H^2+\dot{H}\)+4H\dot{\bar{f}}_{QQ}\]H\nabla^2\dot{E} \nonumber \\
&-\dfrac{1}{3}\(\bar{f}_{Q}+12\bar{f}_{QQ}H^2\)\nabla^2\ddot{E}.
\end{align}
By taking the perturbation of the continuity equation \eqref{9}, the i- components of it becomes
\begin{align}\label{46}
&\(\bar{\rho}^{[\phi]}+\bar{p}^{[\phi]}\)\(\kappa+\bar{f}_{T}\)a\delta u \nonumber\\
&=2\(\bar{f}_{Q}+3\dot{H}\bar{f}_{QQ}\)H\Psi
+18H\dot{H}\bar{f}_{QQ}\Phi \nonumber \\
&+2\(\bar{f}_{Q}+6H^2\bar{f}_{QQ}\)\dot{\Phi}
-6H\dot{H}\bar{f}_{QQ}\nabla^2 E \nonumber \\
&-4H^2\bar{f}_{QQ}\nabla^2 \dot{E}
+4H^2\bar{f}_{QQ}a^{-1}\nabla^2 W,
\end{align}
and the 0-component of them can be written in the following form  
\begin{align}\label{47}
&\[\kappa+\bar{f}_{T}+{1\over2}\bar{f}_{T}\(1-\dfrac{\delta\dot{p}^{[\phi]}}{\delta\dot{\rho}^{[\phi]}}\)\]\dot{\delta} \nonumber\\
&-\Bigg\lbrace \(\kappa+\bar{f}_{T}\)\[{2\(\kappa+\bar{f}_{T}\)\bar{\omega}-\bar{f}_{T}\(1-c_s^2\)\over 2(\kappa+\bar{f}_{T})+\bar{f}_{T}\(1-c_s^2\)}-{\delta p^{[\phi]}\over\delta\rho^{[\phi]}}\] \nonumber \\
&+{\(\kappa+\bar{f}_{T}\)\(\bar{\omega}+1\)\over
 2(\kappa+\bar{f}_{T})+\bar{f}_{T}\(1-c_s^2\)} \nonumber \\
 &\times\Bigg[ \bar{f}_{T}\(1-{\delta\dot{p}^{[\phi]}\over\delta\dot{\rho}^{[\phi]}}\)
+\bar{f}_{TT}\bar{\rho}^{[\phi]}\(1-3{\delta p^{[\phi]}\over\delta\rho^{[\phi]}}\)\(1-3c_s^2\) \Bigg]\nonumber \\
&+\bar{f}_{TT}\bar{\rho}^{[\phi]}(\bar{\omega}+1)\Big(1-3{\delta
p^{[\phi]}\over\delta\rho^{[\phi]}}\Big)\Bigg[{\bar{f}_{T}\(1-c_s^2\)\over 2\kappa+\bar{f}_{T}\(3-c_s^2\)}\Bigg]\Bigg\rbrace 3H\delta \nonumber\\
&+\(\kappa+\bar{f}_{T}\)\(\bar{\omega}+1\)
\bigg[a^{-1}\nabla^2\(\delta u-W\)-3\dot{\Phi}+\nabla^2\dot{E}\bigg]\nonumber \\
&=\dfrac{\delta B_0}{a\bar{\rho}^{[\phi]}}.
\end{align}
We would like to emphasize that the right-hand side of this equation, $\delta B_0$, arises from the perturbation of the zeroth component of $B_\mu$, which is a result of the presence of non-metricity.
\begin{align}\label{48}
&\delta B_0=-a^{-1}\nabla^2\Bigg[\(8H\bar{f}_Q+{1\over2}\dot{\bar{f}}_{Q}\)\Phi-12H^2\bar{f}_{QQ} \dot{\Phi} \nonumber\\
&-\({1\over2}\dot{\bar{f}}_Q+12H^3\bar{f}_{QQ}\)\Psi
+{1\over2}\(16H^2\bar{f}_Q+7H\dot{\bar{f}}_{Q}+\ddot{\bar{f}}_Q\)aW  \nonumber \\
&-{1\over2}\dot{\bar{f}}_Qa\dot{W}-\({8\over3}H\bar{f}_Q-{1\over2}\dot{\bar{f}}_{Q}\)\nabla^2E+4H^2\bar{f}_{QQ}\nabla^2\dot{E}\nonumber \\
&-2H^2\bar{f}_{QQ}a^{-1}\nabla^2W\Bigg]
\end{align}
It is clear that equations \eqref{44}-\eqref{48} reduce to the equivalent equations in General Relativity \cite{Weinberg} when $\bar{f}_Q=-1$ and $\bar{f}_T=0$.
Using the matter Lagrangian \eqref{18}, we can obtain the first-order perturbations in the pressure and energy density of the scalar field as follows:
\begin{align}
& \label{49} \delta\rho^{[\phi]}= \dot{\bar{\phi}}\,\delta\dot{\phi}
+V'\(\bar{\phi}\)\delta\phi-\dot{\bar{\phi}}^{2}\Psi, \\
& \label{50} \delta p^{[\phi]}= \dot{\bar{\phi}}\,\delta\dot{\phi}
-V'\(\bar{\phi}\)\delta\phi-\dot{\bar{\phi}}^{2}\Psi,
\end{align}
Also, the perturbation to the scalar velocity potential $\delta u$ is given by $\delta u=-{\delta \phi/\dot{\bar{\phi}}}$. 

\subsection{Scalar perturbations in $f(Q,T)=\alpha Q+g(T)$}

We have seen, that there exists a system of coupled differential equations, specifically equations (\ref{44}-\ref{50}), for the unknown perturbed quantities in the general form of $f(Q,T)$ gravity.
In this subsection, we will solve these set of differential equations and examine the evolution of perturbed quantities within the functional $f(Q,T)$ in the form $f(Q,T)=\alpha Q+g(T)$.
As the quantities in the background slowly evolve during inflation, we will utilize slow-roll approximations for all relevant equations in the following discussion. We can rewrite equation \eqref{44} for the case where $f(Q,T)$ takes the form $f(Q,T)=\alpha Q+g(T)$ as follows
\begin{align}\label{53}
&\delta\rho^{[\phi]}\Big[\kappa+{3\over2}\bar{g}_{,T}-\bar{g}_{,TT}\Big(\bar{\rho}^{[\phi]}+\bar{p}^{[\phi]}\Big)\Big] \nonumber\\
&+\delta p^{[\phi]}\Big[-{1\over2}\bar{g}_{,T}+3\bar{g}_{,TT}\(\bar{\rho}^{[\phi]}+\bar{p}^{[\phi]}\)\Big]\nonumber\\
&= 6\alpha H(H\Psi+\dot{\Phi})-2\alpha a^{-2}\nabla^2\Phi
+2\alpha H a^{-1}\nabla^2W.
\end{align} 
Also, in this case, the equation \eqref{45} becomes
\begin{align}\label{54}
&{\dfrac{1}{4}}\bar{g}_{,T}\delta\rho^{[\phi]}-\Big(\frac{\kappa}{2}
+{3\over4}\bar{g}_{,T}\Big)\delta p^{[\phi]}=\alpha\(H\dot{\Psi}+3H\dot{\Phi}+\ddot{\Phi}\) \nonumber \\
&+\alpha\(3H^2+\dot{H}\)\Psi+\dfrac{\alpha}{3}a^{-2}\nabla^2\Psi-\dfrac{\alpha}{3}a^{-2}\nabla^2\Phi \nonumber \\
&+\dfrac{2\alpha}{3}Ha^{-1}\nabla^2W+\dfrac{\alpha}{3}a^{-1}\nabla^2\dot{W}-\alpha H\nabla^2\dot{E}-\dfrac{\alpha}{3}\nabla^2\ddot{E}.
\end{align} 
Additionally, The equation \eqref{46} can be written in the simple form  
\begin{equation}\label{55}
\(\bar{\rho}^{[\phi]}+\bar{p}^{[\phi]}\)(\kappa+\bar{g}_{,T})a\delta u =2\alpha\(H\Psi+\dot{\Phi}\).
\end{equation}  
Moreover, in this instance, the perturbed equation \eqref{47} transforms into
\begin{align}\label{56}
&\[\kappa+\bar{g}_{,T}+{1\over2}\bar{g}_{,T}\(1-\dfrac{\delta\dot{p}^{[\phi]}}{\delta\dot{\rho}^{[\phi]}}\)\]\dot{\delta} \nonumber\\
&-\Bigg\lbrace \(\kappa+\bar{g}_{,T}\)\[{2\(\kappa+\bar{g}_{,T}\)\bar{\omega}-\bar{g}_{,T}\(1-c_s^2\)\over 2\(\kappa+\bar{g}_{,T}\)+\bar{g}_{,T}\(1-c_s^2\)}-{\delta p^{[\phi]}\over\delta\rho^{[\phi]}}\] \nonumber \\
&+{\(\kappa+\bar{g}_{,T}\)\(\bar{\omega}+1\)\over
 2(\kappa+\bar{g}_{,T})+\bar{g}_{,T}\(1-c_s^2\)}
 \Bigg[\bar{g}_{,T}\(1-{\delta\dot{p}^{[\phi]}\over\delta\dot{\rho}^{[\phi]}}\) \nonumber \\
&+\bar{g}_{,TT}\bar{\rho}^{[\phi]}\(1-3{\delta p^{[\phi]}\over\delta\rho^{[\phi]}}\)\(1-3c_s^2\) \Bigg]\nonumber \\
&+\bar{g}_{,TT}\bar{\rho}^{[\phi]}(\bar{\omega}+1)\Big(1-3{\delta
p^{[\phi]}\over\delta\rho^{[\phi]}}\Big)\Bigg[{\bar{g}_{,T}\(1-c_s^2\)\over 2\kappa+\bar{g}_{,T}\(3-c_s^2\)}\Bigg]\Bigg\rbrace3H \delta \nonumber\\
&+\(\kappa+\bar{g}_{,T}\)\(\bar{\omega}+1\)
\bigg[a^{-1}\nabla^2\(\delta u-W\)-3\dot{\Phi}+\nabla^2\dot{E}\bigg] \nonumber\\
&=\dfrac{\delta B_0}{a\bar{\rho}^{[\phi]}},
\end{align} 
where in this relation $\delta B_0$ from relation \eqref{48} can be written as
\begin{equation}\label{57}
\delta B_0=-8\alpha\( Ha^{-1}\nabla^2\Phi-H^2a\nabla^2W-{1\over3}H\nabla^4E\).
\end{equation}
From these four equations, we need to derive the perturbed quantities $\delta \phi$ and $\Phi$. However, using an alternative gauge greatly simplifies the solution. Consequently, we proceed with the calculations using an different gauge, based on the assumption that \cite{Weinberg}  
\begin{equation}\label{58}
E=0,~~~\delta\phi=0.
\end{equation}
In this gauge, the perturbations in pressure, energy density and velocity derived from equations \eqref{49} and \eqref{50} are expressed as follows: 
\begin{equation}\label{58.1}
\delta\rho=\delta p={1\over2}h_{00}\dot{\bar{\phi}}^2=\dot{\bar{\phi}}^2\Psi,~~~~~\delta u=0.
\end{equation}
By substituting $\dot{\bar{\phi}}^2$ from equation \eqref{22} into equation \eqref{58.1} we obtain 
\begin{equation}\label{59}
\delta\rho=\delta p=-\dfrac{2\alpha}{\kappa+g_T}\dot{H}\Psi.
\end{equation}
Also, in this gauge, equations \eqref{53} and \eqref{55} reduces to
\begin{align}\label{60}
-\Big[{\(\kappa+\bar{g}_{,T}\)^2+4\alpha\bar{g}_{,TT}\dot{H}\over\(\kappa+\bar{g}_{,T}\)^2}\Big]\dot{H}\Psi
&=3 H(H\Psi+\dot{\Phi}) \nonumber \\
-a^{-2}\nabla^2\Phi
+H a^{-1}\nabla^2W,\\ 
\label{61}
H\Psi+\dot{\Phi}&=0,
\end{align}
and the equation \eqref{56} becomes 
\begin{align}\label{62}
&\dot{\delta}-\Bigg\lbrace \[{2\(\kappa+\bar{g}_{,T}\)\bar{\omega}-\bar{g}_{,T}\(1-c_s^2\)\over 2\(\kappa+\bar{g}_{,T}\)+\bar{g}_{,T}\(1-c_s^2\)}-1\]
 \nonumber \\
&-2\bar{g}_{,TT}\bar{\rho}^{[\phi]}(\bar{\omega}+1)\Bigg[{\kappa(1-3c_s^2)+2\bar{g}_{,T}\(1-2c_s^2\)\over \(2\kappa
+\bar{g}_{,T}\(3-c_s^2\)\)(\kappa+\bar{g}_{,T})}\Bigg]\Bigg\rbrace 3H\delta \nonumber\\
&-\(\bar{\omega}+1\)a^{-1}\nabla^2W-3\(\bar{\omega}+1\)\dot{\Phi}
=\dfrac{\delta B_0}{a\bar{\rho}^{[\phi]}}.
\end{align} 
By substituting of $\delta=\delta\rho^{[\phi]}/\bar{\rho}^{[\phi]}$ and 
 $\dot{\delta}=\dot{\delta\rho}^{[\phi]}/\bar{\rho}^{[\phi]}-\dot{\bar{\rho}}^{[\phi]}\delta\rho^{[\phi]}/{\rho^{[\phi]}}^2$ into the equation \eqref{62} we obtain
\begin{align}\label{63}
&\dot{\delta\rho}^{[\phi]}-\Bigg\lbrace\[{\kappa\(\bar{\omega}-1\)+\bar{g}_{,T}\(\bar{\omega}-2+c_s^2\)\over \(2\kappa+\bar{g}_{,T}\(3-c_s^2\)\)}\]
 \nonumber \\
&-2\bar{g}_{,TT}\bar{\rho}^{[\phi]}(\bar{\omega}+1)\Bigg[{\kappa(1-3c_s^2)+2\bar{g}_{,T}\(1-2c_s^2\)\over \(2\kappa
+\bar{g}_{,T}\(3-c_s^2\)\)(\kappa+\bar{g}_{,T})}\Bigg] \nonumber \\
&-{\dot{\bar{\rho}}^{[\phi]}\over3H\bar{\rho}^{[\phi]}} \Bigg\rbrace 3H\delta\rho^{[\phi]} \nonumber\\
&-\(\bar{\omega}+1\)a^{-1}\nabla^2W-3\(\bar{\omega}+1\)\dot{\Phi}
=\dfrac{\delta B_0}{a\bar{\rho}^{[\phi]}},
\end{align} 
We need an equation that characterizes the evolution of the perturbed quantity $\Phi$. To achieve this, we eliminate $\Psi$ from equation \eqref{63} by applying relation \eqref{61} as follows:
\begin{align}\label{64}
&\ddot{\Phi}+\(-3H+{\ddot{H}\over \dot{H}}-{\dot{H}\over H}
-{\dot{\bar{g}}_{,T}\over (\kappa+\bar{g}_{,T})}\)\dot{\Phi} \nonumber \\
&-\Bigg\lbrace\({\kappa\(\bar{\omega}-1\)+\bar{g}_{,T}\(\bar{\omega}-2+c_s^2\)\over \(2\kappa+\bar{g}_{,T}\(3-c_s^2\)\)}\)
 \nonumber \\
&-\bar{g}_{,TT}\bar{\rho}^{[\phi]}(\bar{\omega}+1)\({\kappa(1-3c_s^2)+2\bar{g}_{,T}\(1-2c_s^2\)\over \(2\kappa
+\bar{g}_{,T}\(3-c_s^2\)\)(\kappa+\bar{g}_{,T})}\)\nonumber \\
&-{\dot{\bar{\rho}}^{[\phi]}\over6H\bar{\rho}^{[\phi]}} \Bigg\rbrace 6H\dot{\Phi} 
-Ha^{-1}\nabla^2W=0.
\end{align}
We can obtain the quantity $Ha^{-1}\nabla^2W$ using equation \eqref{60}, and then substitute this back into \eqref{64}, leading to:
\begin{align}\label{65}
&\ddot{\Phi}+\Bigg\lbrace-3H+{\ddot{H}\over \dot{H}}-2{\dot{H}\over H}
-{\dot{\bar{g}}_{,T}\over (\kappa+\bar{g}_{,T})} \nonumber\\
&-6H\({\kappa\(\bar{\omega}-1\)+\bar{g}_{,T}\(\bar{\omega}-2+c_s^2\)\over \(2\kappa+\bar{g}_{,T}\(3-c_s^2\)\)}\)
 \nonumber \\
&+12\alpha\bar{g}_{,TT}H\dot{H}\({\kappa(1-3c_s^2)+2\bar{g}_{,T}\(1-2c_s^2\)\over \(2\kappa+\bar{g}_{,T}\(3-c_s^2\)\)\(\kappa+\bar{g}_{,T}\)^2}+{\dot{H}\over3H^2}\) \nonumber \\
&-{\dot{\bar{\rho}}^{[\phi]}\over\bar{\rho}^{[\phi]}} \Bigg\rbrace \dot{\Phi} 
-a^{-2}\nabla^2\Phi=0,
\end{align}
We derived equation \eqref{65} without employing any approximations. In the following discussion, we will utilize the slow-roll approximation for the background quantities. As a result, at the leading order of the slow-roll parameters, we obtain
 \begin{align}\label{66}
&{\kappa\(\bar{\omega}-1\)+\bar{g}_{,T}\(\bar{\omega}-2+c_s^2\)\over \(2\kappa+\bar{g}_{,T}\(3-c_s^2\)\)}\approx-1+{\kappa+\bar{g}_{,T}\over\kappa+2\bar{g}_{,T}}\eta, \\
\label{66.1}
&{\kappa(1-3c_s^2)+2\bar{g}_{,T}\(1-2c_s^2\)\over \(2\kappa+\bar{g}_{,T}\(3-c_s^2\)\)\(\kappa+\bar{g}_{,T}\)^2}\approx {2\kappa+3\bar{g}_{,T}\over \kappa+2\bar{g}_{,T}} \nonumber \\
&+{3\kappa\(2\kappa+3\bar{g}_{,T}\)+\bar{g}_{,T}\(2\kappa+5\bar{g}_{,T}\)\over\(\kappa+2\bar{g}_{,T}\)\(\kappa+\bar{g}_{,T}\)}\lambda, \\
\label{66.2}
&{\dot{\bar{\rho}}^{[\phi]}\over\bar{\rho}^{[\phi]}}\approx-6H{\kappa+\bar{g}_{,T}\over \kappa+2\bar{g}_{,T}}\eta.
\end{align}
Finally, by substituting equations (\ref{66}-\ref{66.2}) into equation \eqref{65} and performing straightforward calculations, we arrive at
\begin{align}\label{67}
&\ddot{\Phi}+\Bigg\lbrace3H+{\ddot{H}\over \dot{H}}-2{\dot{H}\over H}
+{\bar{g}_{,T}\over 2(\kappa+\bar{g}_{,T})}{\dot{\bar{g}}_{,T}\over (\kappa+\bar{g}_{,T})} \Bigg\rbrace \dot{\Phi} \nonumber\\
&-a^{-2}\nabla^2\Phi=0.
\end{align}
In order to construct quantities that are gauge invariant up to first order in slow roll parameters, one can take specific combinations of the perturbation quantity such that they don't change under an infinitesimal diffeomorphism.
For the scalar perturbations, one can choose the following gauge invariant combination:
\begin{equation}\label{68}
\mathcal{R}=-\Phi+H\delta u.
\end{equation}
In general, the gauge invariant variable $\mathcal{R}$ can be written as superpositions of these two solutions 
\begin{equation}\label{69}
\mathcal{R}({\bf x},t)=\int d^3k\[\mathcal{R}_k(t)e^{i{\bf k}.{\bf x}}+\mathcal{R}^{\star}_k(t)e^{-i{\bf k}.{\bf x}}\].
\end{equation} 
In this gauge, as indicated by equation \eqref{58.1}, we have $\mathcal{R}=-\Phi$. Consequently, the equation \eqref{67} that characterizes the evolution of $\Phi$ is also applicable to $\mathcal{R}$.
Thus, we can express equation \eqref{67} in terms of the conformal time $\tau$, defined by the relation $d\tau=a^{-1} dt$, for the gauge-invariant quantity $\mathcal{R}_k$ as follows:
\begin{equation}\label{70}
{d^2\mathcal{R}_k\over d\tau^2}+\Big({2\over z}{dz\over d\tau}\Big) {d\mathcal{R}_k\over d\tau}
+k^{2}\mathcal{R}_k=0,
\end{equation}
where in this relation $dz/z d\tau$ at leading order is  
\begin{equation}\label{71}
{1\over z}{dz\over d\tau}=aH\(1+\delta+\epsilon+{3\bar{g}_{,T}\over4\(\kappa+\bar{g}_{,T}\)}\theta\).
\end{equation}
This is the Mukhanove-Sasaki equation, which has been derived in the framework of $f(Q,T)$ gravity in the functional form $f(Q,T) = \alpha Q + g(T)$, using the slow roll approximation.
By solving equation \eqref{71}, one can express $z$ in the following manner
\begin{align}\label{72}
&z\equiv\sqrt{{a^2\dot{H}\over H^2}\sqrt{U}},\\
&U=C\(\kappa+\bar{g}_{,T}\)\exp\({\kappa\over \kappa+\bar{g}_{,T}}\),
\end{align}
where $C$ is constant of integration. We can determine the value of $C$ such that $U$ equals one when $g_{,T}=\beta$. Thus, we can express $C$ as $C= (\kappa+\beta)^{-1} \exp\left(-\frac{\kappa}{\kappa+\beta}\right)$.
Also, by substituting $C$ into equation \eqref{72}, we can write $U$ in the following form
\begin{equation}\label{72.1}
U=\({\kappa+\bar{g}_{,T}\over \kappa+\beta}\)\exp\({\kappa(\beta-\bar{g}_{,T})\over (\kappa+\beta)(\kappa+\bar{g}_{,T})}\).
\end{equation}
 Notice that in the investigation of fluctuations, we can assume that the slow roll parameters are nearly constant during inflation, therefore integrating 
${d\over d\tau}({1\over aH})=-1+\epsilon$ then gives $aH=-1/(1-\epsilon)\tau$. 
Thus, by introducing canonically-normalized Mukhanove variable $v_k=z\mathcal{R}_k$, equation \eqref{70} can be rewritten in the well-known format of the Mukhanov-Sasaki equation as follows  
\begin{equation}\label{73}
{d^2v_k\over d\tau^2}+\(k^{2}-{1\over z}{d^2z\over d\tau^2}\)v_k=0,
\end{equation}
where in this relation we have
\begin{eqnarray}\label{74} 
{1\over z}{d^2z\over d\tau^2}&=&{1\over \tau^2}(\nu^2-{1\over4}),\\
\label{74.1}
\nu&=&{3\over2}+\delta+2\epsilon+{3\bar{g}_{,T}\over4\(\kappa+\bar{g}_{,T}\)}\theta.
\end{eqnarray}
For constant and real $\nu$, the general solution to equation \eqref{73} is given by
\begin{equation}\label{75}
v_k(\tau)=\sqrt{-\tau}\[C_1 H^{(1)}_\nu(-k\tau)+C_2 H^{(2)}_\nu(-k\tau)\],
\end{equation}
where $H^{(1)}_\nu$ and $H^{(2)}_\nu$ are the Hankel's function of the first and second kind, respectively. 
Also, from equation \eqref{74} one can conclude that $z=\tau^{1/2-\nu}$. 
We will now set the boundary conditions by applying the Bunch-Davies vacuum, given by $v_k \rightarrow \frac{e^{-ik\tau}}{\sqrt{2k}}$ in the ultraviolet regime where $-k\tau \gg 1$. This leads to the results $C_1=\frac{\sqrt{\pi}}{2}\exp\left(i\frac{\pi}{2}\left(\nu+\frac{1}{2}\right)\right)$ and $C_2=0$. Consequently, we can express the solution \eqref{75} in the following form:
\begin{equation}\label{76}
v_k(\tau)=-{\sqrt{-\pi\tau}\over2}e^{i{\pi\over2}(\nu+{1\over2})} H^{(1)}_\nu(-k\tau).
\end{equation}
Additionally, on super-horizon scales where $-k\tau \ll 1$, equation \eqref{76} approaches the asymptotic value of
\begin{equation}\label{77}
v_k(\tau)=-i{2^{\nu-2}\over\sqrt{k}}{\Gamma(\nu)\over\Gamma\({3/2}\)}e^{i{\pi\over2}(\nu+{1/2})}\({-k\tau}\)^{1/2-\nu}.
\end{equation}
Ultimately, the scalar power spectrum of curvature fluctuations can be expressed as 
\begin{eqnarray}\label{78}
\mathcal{P}_s={k^3\over 2\pi^2}\vert\mathcal{R}_k\vert^2&\approx&{1\over8\pi^2}\({H_{\star}^4\over {\dot{H}}_{\star}\sqrt{U_\star}}\)  \nonumber \\
&\approx&{\alpha\over2\pi^2}\({H^4\over\dot{\phi}^2\(\kappa+\bar{g}_{,T}\)\sqrt{U}}\)\Bigg\vert_{k=aH}
\end{eqnarray}
where the index $\star$ for any quantity indicates that this quantity is evaluated at the horizon crossing $k=aH$. 
Moreover the spectral index is given by
\begin{eqnarray}\label{79}
n_s-1&=&{d Ln(\mathcal{P}_s)\over dLnk}\Bigg\vert_{k=aH}=-4\epsilon-6\lambda-{3\over2}\({2\kappa+3\bar{g}_{,T}\over\kappa+\bar{g}_{,T}}\)\theta \nonumber \\
&=& -4\epsilon+2\delta-{3\over2}\({\bar{g}_{,T}\over\kappa+\bar{g}_{,T}}\)\theta.
\end{eqnarray}
As we have noted, in the spatial scenarios where $\bar{g}_{,T} = 0$ or in the linear form $f(Q,T)$ gravity, we find that $\theta = 0$, and the spectral index becomes $n_s=1-4\epsilon+2\delta$, which corresponds to the spectral index found in general relativity.

\subsection{Tensor perturbations}

In this subsection, we will examine the evolution of tensor perturbations. We introduce these tensor perturbations into the metric in the following manner:
\begin{equation}\label{80}
ds^2=-dt^2+a^2(t)(\delta_{ij}+h_{ij}) dx^i dx^j,
\end{equation}
where tensor perturbations are characterized by the tensor $h_{ij}$, with properties such that $h_{ij}=h_{ji}$, $\partial_ih_{ij}=0$ and $h_{ii}=0$. By perturbing the energy-momentum tensor up to the second-order within the framework of linear perturbations as \cite{Antonio2022}
\begin{equation}\label{81}
\ddot{h}_{ij}+\Big\lbrace
3H-\dfrac{d}{dt}\Big[\ln(f_Q)\Big]\Big\rbrace
\dot{h}_{ij}-\dfrac{\nabla^2}{a^2} h_{ij}=0.
\end{equation}
In our proposed model, characterized by the functional form $f(Q,T)=\alpha Q+g(T)$ in the context of $f(Q,T)$ gravity, the third term in relation \eqref{81} becomes zero, leading to $\frac{d}{dt}[\ln(f_Q)]=0$.
Therefore, equation \eqref{81} is simplified to an evolution equation for
tensor perturbations in minimal coupling models of general relativity. 
By substituting the plan wave solutions of the form $h_k(t)e^{i{\bf k}.{\bf x}}e_{ij}$ where $e_{ij}$ is a time independent polarization tensor, into equation \eqref{81} we derive the following result 
\begin{equation}\label{82}
\ddot{h}_{k}+3H\dot{h}_{k}+\dfrac{k^2}{a^2} h_{k}=0.
\end{equation}
Also, in conformal time, equation \eqref{82} becomes
\begin{equation}\label{83}
{d^2h_{k}\over d\tau^2}+2{dLnz_t\over d\tau} {d{h}_{k}\over d\tau}
+k^2 h_{k}=0,
\end{equation}
where $z_t=a$ and $z^{''}/z=a^2H^2(2-\epsilon)=\tau^{-2}(2+3\epsilon)$. By substitution the canonical variable $u_k=z_th_k$ for tensors perturbations, into equation \eqref{83} we obtain
\begin{equation}\label{84}
{d^2u_{k}\over d\tau^2}+\[k^2-{1\over\tau^2}\(\mu^2-{1\over4}\)\]{u}_{k}=0,
\end{equation} 
where $\mu\approx3/2+\epsilon$. 
The solution of this differential equation is the Hankel function of the first kind and second kind. By imposing the Bunch-Davies vacuum in the ultraviolet regime, analogous to the method used for scalar perturbations, we derive     
\begin{equation}\label{85}
\mathcal{P}_t={4H^2\over \pi^2}.
\end{equation} 
Consequently, the tensor primordial power spectrum remains invariant, similar to its state during slow-roll inflation in the context of general relativity, without any modifications.
Furthermore, tensor-to-scalar ratio $r$, from relations \eqref{78} and \eqref{85} is expressed as
\begin{equation}\label{86}
r=\dfrac{\mathcal{P}_t}{\mathcal{P}_s}\approx16\epsilon\sqrt{U}.
\end{equation}
We calculate the scalar power spectrum $\mathcal{P}_s$, the tensor power spectrum $\mathcal{P}_t$, the spectral index $n_s$, and the tensor-to-scalar ratio $r$ within the framework of $f(Q,T)$ gravity, specifically in the functional form $f(Q,T) = \alpha Q + g(T)$, where $g(T)$ represents an arbitrary function of $T$. The noteworthy aspect is that when $g(T) = 0$, all of this quantity simplifies to a form that is consistent with general relativity.
The significant point is that when $g(T) = 0$, the entire expression simplifies to a form consistent with the principles of general relativity.

\begin{figure*}[t]
\begin{center}
 \subfloat{\includegraphics[width=7cm]{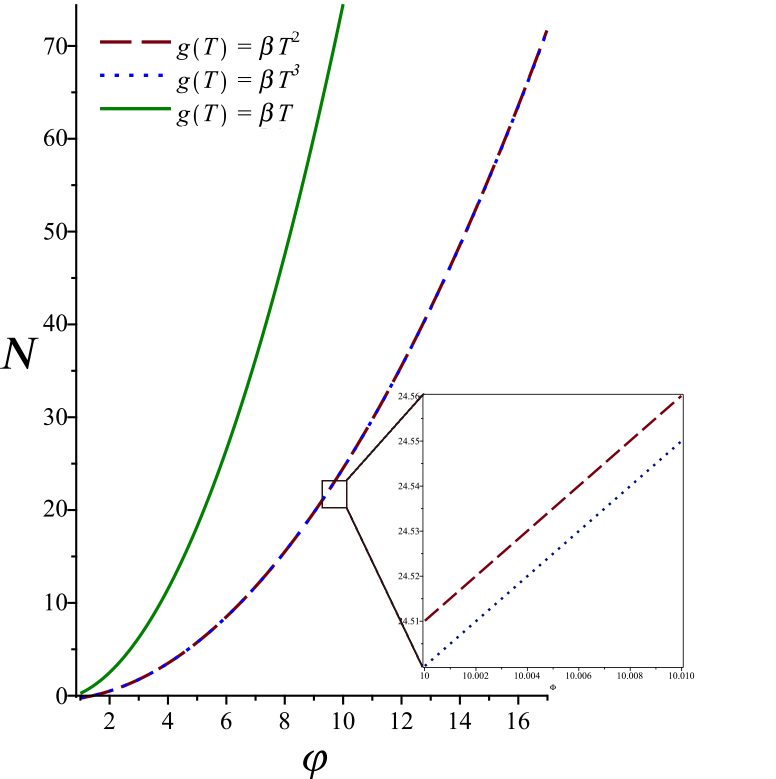}}
   \subfloat{\includegraphics[width=7cm]{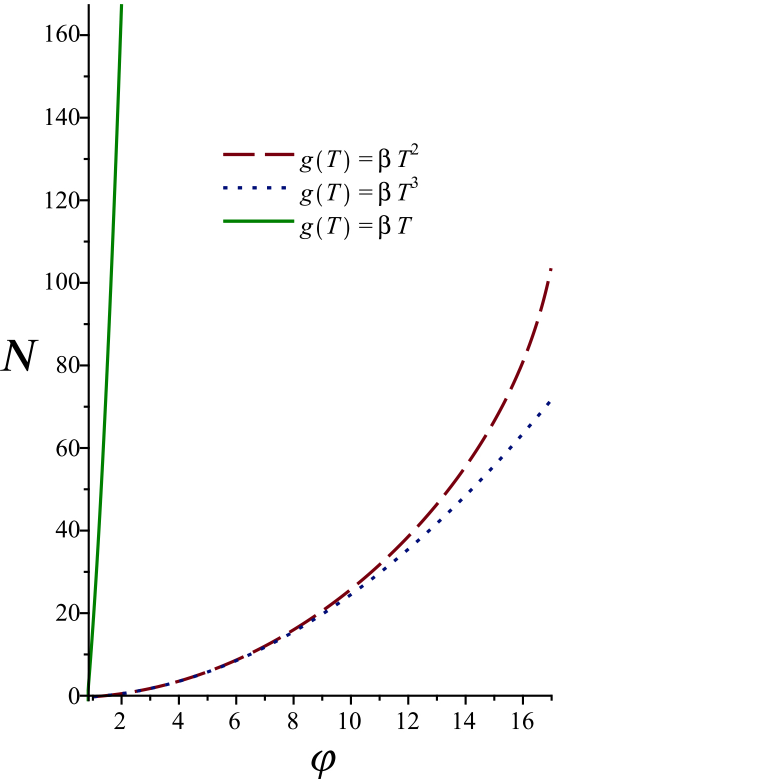}}
              \end{center}
 \caption{\footnotesize (color online) The relationship between the number of e-folds $\mathcal{N}$ versus the dimensionless scalar field $\varphi=\kappa\phi$ is depicted for a quadratic potential $V(\phi)=M\phi^2$ and three forms of the function $g(T):$ $\beta T$, $\beta T^2$ and $\beta T^3$. (left) shows the case with  $\alpha=-1$, $\beta=2$, $M=10^{-6}$, while  (right) shows the case with $\alpha=-1$, $\beta=200$ and $M=10^{-6}$.}
\label{fig1}
\end{figure*}

\section{Application to a specific model of inflation}

In the Previous sections, we obtained the slow roll parameters, power spectrum, spectral index and tensor-to-scalar ratio in $f(Q,T)$ gravity, specifically in the functional form $f(Q,T)=\alpha Q + g(T)$.
We will now investigate specific cosmological models within the context of $f(Q,T)$ gravity, concentrating on different choices for the functional form of $f(Q,T) = \alpha Q + g(T)$. Our analysis will include the simple forms $f(Q,T) = \alpha Q + \beta T$ and $f(Q,T) = \alpha Q + \beta T^2$. Additionally, we will assess our findings against observational data using a quadratic potential given by $V(\phi) = M^2\phi^2$.
Up to this point, we have denoted background quantities with an overbar; for instance,$\bar{g}=g(\bar{T})$. Because we are working to linear order, we can substitute $\bar{g}$ with $\tilde{g}$ when $\bar{g}$ or any function of it is multiplied by a slow-roll parameter.

\subsection{$f(Q,T)=\alpha Q+\beta T$}
As a simple example of the cosmological inflation in $f(Q,T)$ gravity we will examine the case in which function $f(Q,T)$ has a linear form, $f(Q,T)=\alpha Q+\beta T$, where $\alpha$ and $\beta$ are constant.
So we simply obtain $g(T)=\beta T$, $\tilde{g}=-4\beta V(\bar{\phi})$ and $\tilde{g}_{,T}=\beta$. Thus the slow roll parameters from equations (\ref{31}-\ref{35}) become
\begin{align}\label{87}
&\lambda\approx\dfrac{\alpha}{6(\kappa+\beta)}\({2V V^{''}-{V^{'}}^2\over V^2}\),\\
&\eta\approx-\dfrac{\kappa+2\beta}{3(\kappa+\beta)}\epsilon,\\
&\theta=0 \\
&\epsilon\approx-\dfrac{\alpha}{2(\kappa+\beta)}\({V^{'}\over V}\)^2,\\
&\delta\approx-3\lambda.
\end{align}
Also, from equation \eqref{37}, the e-folding number is given by
\begin{equation}\label{88}
\mathcal{N}\approx{\kappa+\beta\over \alpha}\int_{\phi_{\star}}^{\phi_{end}}
\dfrac{V}{V^{\prime}}{d}\phi.
\end{equation}
The power spectrum, spectral index and the tensor to scalar ratio in this framework is given by
\begin{eqnarray}\label{89}
\mathcal{P}_s&\approx&{\alpha\over2\pi^2\(\kappa+\beta\)}\({H_{\star}^4\over\dot{\phi}_{\star}^2}\),\\
n_s-1&\approx&-4\epsilon-6\lambda=-4\epsilon+2\delta\\
r&\approx&16\epsilon.
\end{eqnarray}
As an example, we will examine the power-law potential given by $V(\phi)=M\phi^n$, where $M$ and $n$ are positive constants \cite{Linde1983} within the framework of linear $f(Q,T)$ gravity. So,the slow-roll parameters becomes
\begin{equation}\label{90}
\lambda\approx\dfrac{\alpha n(n-2)}{6(\kappa+\beta)}{1\over \phi^2},~~
\epsilon\approx-\dfrac{\alpha n^2}{2(\kappa+\beta)}{1\over \phi^2},~~
\theta=0,
\end{equation}
and the number of e-folds is given by 
\begin{equation}\label{91}
\mathcal{N}\approx-\dfrac{(\kappa+\beta)}{2\alpha n}(\phi_{\star}^2-\phi_{end}^2).
\end{equation}
We can derive $\phi_{end}^2=-\alpha n^2/2(\kappa+\beta)$ by setting $\epsilon=1$. 
Moreover, the scalar spectral index and the tensor to scalar ratio can be written as
\begin{equation}\label{92}
n_s-1\approx\dfrac{\alpha n(n+2)}{(\kappa+\beta)}{1\over\phi_{\star}^2},~~~
r\approx-\dfrac{8\alpha n^2}{(\kappa+\beta)}{1\over \phi_{\star}^2},
\end{equation}
We can determine the spectral index and the tensor-to-scalar ratio as a function of the number of e-folds $\mathcal{N}$ by assuming that $\phi_{end}^2 \ll \phi_{\star}^2$ as follow
\begin{equation}\label{93}
n_s-1\approx-\dfrac{(n+2)}{2\mathcal{N}},~~~~
r\approx\dfrac{4n}{\mathcal{N}}.
\end{equation}
We have noted that the observational quantities $n_s$ and $r$ do not depend on the model parameters $\alpha$ and $\beta$. This scenario is completely similar to what is encountered in general relativity with a minimally coupled scalar field.

\begin{figure*}[t]
\begin{center}
 \subfloat{\includegraphics[width=18cm]{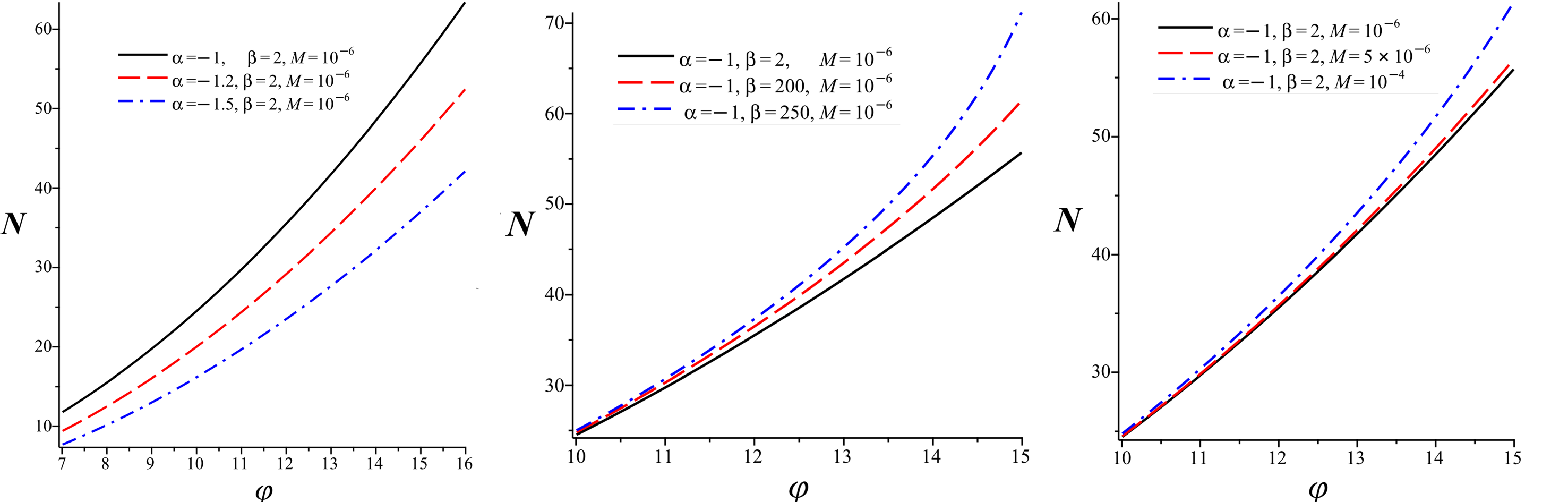}}
            \end{center}
   \caption{\footnotesize (color online) The number of e-folds $\mathcal{N}$ versus the dimensionless scalar field $\varphi=\kappa\phi$ for the quadratic potential $V(\phi)=M\phi^2$ and a function $g(T)=\beta T^2$ has been depicted. Parameter variations are shown: (left) $\beta=2$, $M=10^{-6 }$, and several different values of the parameter $\alpha $; (middle) $\alpha=-1$, $M=10^{-6 }$, and several different values of the parameter $\beta $; (right) $\alpha=-1$, $\beta=2 $, and several different values of the parameter $M$.}
\label{fig2}
\end{figure*}

\subsection{$f(Q,T)=\alpha Q+\beta T^2$}

As a second example of the cosmological inflation in $f(Q,T)$ gravity we will study the case in which function $f(Q,T)$ is given by $f(Q,T)=\alpha Q+\beta T^2$, where $\alpha$ and $\beta$ are constant.
So we simply obtain $g(T)=\beta T^2$, $\tilde{g}=16\beta V^2$ and $\tilde{g}_{,T}=-8\beta V(\bar{\phi})$. 
Thus the slow roll parameters from equations (\ref{31}-\ref{35}) becomes
\begin{eqnarray}\label{94}
\epsilon&\approx&-\dfrac{\alpha\(\kappa-16\beta V\)^2{V^{'}}^2}{2(\kappa-8\beta V)^3V^2}, \\
\theta&\approx&-\dfrac{8\alpha\beta\(\kappa-16\beta V\){V^{'}}^2}{3(\kappa-8\beta V)^3V}, \\
\lambda&\approx&{\kappa\over (\kappa-16\beta V)}\theta+\dfrac{\alpha\(\kappa-16\beta V\)}{3(\kappa-8\beta V)^2}{V^{''}\over V}+{\epsilon\over3}, \\
\label{94.1}
\eta&\approx &{\epsilon\over 3},~~~~
\delta\approx-3\lambda-{3\over2}\theta.
\end{eqnarray}
Also, in this context the number of e-folds is given by
\begin{equation}\label{95}
\mathcal{N}\approx{1\over\alpha}\int_{\phi_{\star}}^{\phi_{end}}\dfrac{(\kappa-8\beta V)^2V}{(\kappa-16\beta V)V^{\prime}}d\phi.
\end{equation}
The power spectrum, spectral index and the tensor to scalar ratio in the context of $f(Q,T)=\alpha Q+\beta T^2$ becomes
\begin{align}\label{96}
&\mathcal{P}_s\approx{\alpha\sqrt{\kappa+\beta}\over2\pi^2\(\kappa-8\beta V\)^{3/2}}{H_{\star}^4\over\dot{\phi}_{\star}^2}\exp\({-\kappa\beta(1+8V)\over2(\kappa+\beta)(\kappa-8\beta V)}\),\\
&n_s-1\approx-4\epsilon+2\delta+\({12\beta V\over\kappa-8\beta V}\)\theta, \\
&r\approx{8\alpha\over\sqrt{\kappa+\beta}}{\(\kappa-16\beta V\)^2{V^{'}}^2\over\(\kappa-8\beta V\)^{5/2}V^2}\exp\({\kappa\beta(1+8V)\over2(\kappa+\beta)(\kappa-8\beta V)}\).
\end{align}
Let's examine the simplest form of inflaton potential, referred to as the chaotic potential, given by $V(\phi) = M\phi^2$, where $M$ is a constant. For this potential, the slow roll parameters from relations \eqref{94}- \eqref{94.1} are given by
\begin{eqnarray}\label{97}
\epsilon&\approx&-{2\alpha\(\kappa-16M\beta\phi^2\)^2\over\phi^2(\kappa-8M\beta\phi^2)^3}, \\
\theta&\approx&-{32\alpha\beta M\(\kappa-16M\beta\phi^2\)\over(\kappa-8M\beta\phi^2)^3},\\
\lambda&\approx&-{16\alpha\beta M\(\kappa+16M\beta\phi^2\)\over3(\kappa-8M\beta\phi^2)^3}.
\end{eqnarray}
Moreover, in this context the number of e-folds can be written as
\begin{align}\label{98}
&\mathcal{N}\approx-{M\beta\over2\alpha}\(\phi_{\star}^4-\phi_{end}^4\)+{3\kappa\over16\alpha}\(\phi_{\star}^2-\phi_{end}^2\)\nonumber \\
&-{\kappa^2\over 256M\alpha\beta}\ln\({16M\beta\phi_{\star}^2-\kappa\over 16M\beta\phi_{end}^2-\kappa}\).
\end{align}
Finally, the spectral index and the tensor to scalar ratio becomes
\begin{align}
\label{99}
&n_s-1\approx \nonumber \\
&-{8\alpha\(1792M^3\beta^3\phi_{\star}^6- 432M^2\beta^2\kappa\phi_{\star}^4+ 32M\beta\kappa^2\phi_{\star}^2- \kappa^3\)\over(\kappa-8M\beta\phi_{\star}^2)^4\phi_{\star}^2}, \\
\label{100}
&r\approx {-32\alpha\(\kappa-16\beta M\phi_{\star}^2\)^2\over\sqrt{\kappa+\beta}\(\kappa-8\beta M\phi_{\star}^2\)^{5/2}\phi_{\star}^2}\nonumber \\
&\times\exp\({-\kappa\beta(1+8M\phi_{\star}^2)\over2(\kappa+\beta)(\kappa-8\beta M \phi_{\star}^2)}\).
\end{align}

\begin{figure*}[t]
\begin{center}
 \subfloat{\includegraphics[width=18cm]{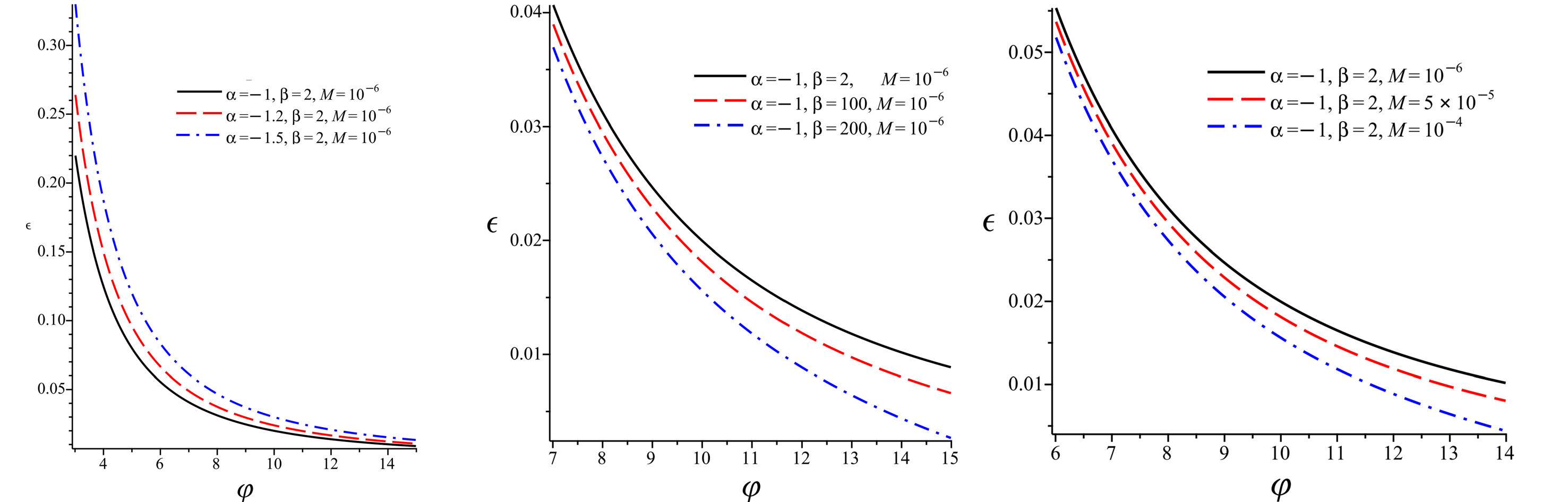}}
                 \end{center}
   \caption{\footnotesize (color online) The slow roll parameter $\epsilon$ versus the dimensionless scalar field $\varphi=\kappa\phi$ for the quadratic potential $V(\phi)=M\phi^2$ and a function $g(T)=\beta T^2$ has been depicted. Parameter variations are shown: (left) $\beta=2$, $M=10^{-6 }$, and several different values of the parameter $\alpha $; (middle) $\alpha=-1$, $M=10^{-6 }$, and several different values of the parameter $\beta $; (right) $\alpha=-1$, $\beta=2 $, and several different values of the parameter $M$.}
\label{fig3}
\end{figure*}

\begin{figure*}[t]
\begin{center}
 \subfloat{\includegraphics[width=9cm]{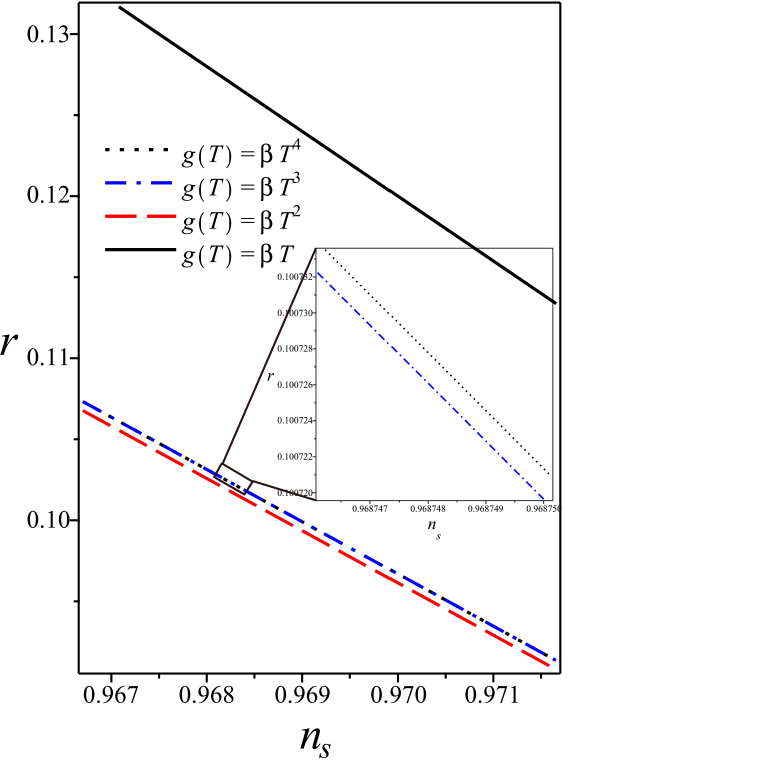}}
   \subfloat{\includegraphics[width=9cm]{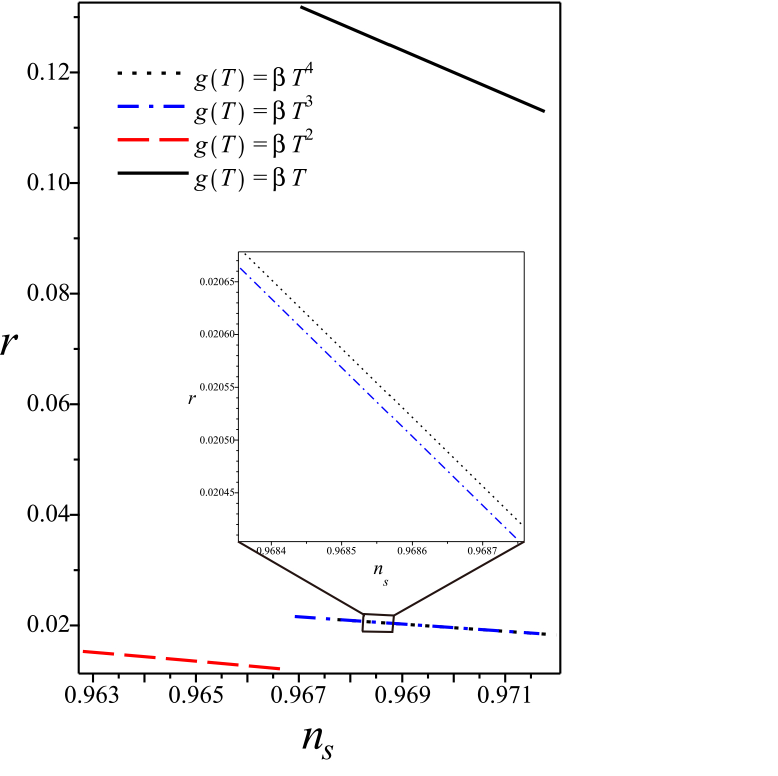}}
              \end{center} 
   \caption{\footnotesize The tensor-to-scalar ratio $r$ versus the scalar spectral index $n_s$ is depicted for a quadratic potential $V(\phi)=M\phi^2$ and four forms of the function $g(T)$: $\beta T$, $\beta T^2$, $\beta T^3$, and $\beta T^4$. The left panel shows the case with $\alpha=-1$, $\beta=2$, $M=10^{-6}$, while the right panel shows the case with $\alpha=-1$, $\beta=200 $, and $M=10^{-6}$. Also for two plots the number of efolds is between $60<N<70$.}
\label{fig4}
\end{figure*}

\begin{figure}[t]
\begin{center}
 \subfloat{\includegraphics[width=9cm]{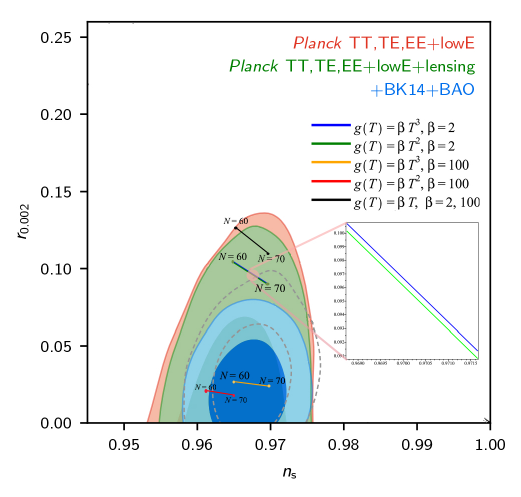}}
  \end{center}
   \caption{\footnotesize (color online) Constraints on the tensor-to-scalar ratio $r_{0.002}$ in the $\Lambda CDM$ model, using Planck TT,TE,EE+lowE and Planck TT,TE,EE+lowE+lensing (red and green, respectively), and joint constraints with BAO and BICEP2/Keck (blue, including Planck polarization to determine the foreground components) \cite{Planck2018}.The tensor-to-scalar ratio $r$ versus the scalar spectral index $n_s$ is depicted for a quadratic potential $V(\phi)=M\phi^2$. This examination considers three forms of the function $ g(T)$: $\beta T$, $\beta T^2$, and $\beta T^3$, as well as two distinct values for $\beta$, specifically $\beta=2$ and $\beta=100$. In this analysis, we also set $\alpha=-1$ and $M=10^{-6} M_p$. Additionally, for all the plots, the number of e-folds falls within the range of $60 < N < 70$. The black line represents the function $g(T)=\beta T$, where the plots for $\beta=2$ and $\beta=100$ overlap perfectly. The blue line represents the function $g(T)=\beta T^3$, where $\beta=2$. The green line represents the function $g(T)=\beta T^2$, where $\beta=2$. The orange line represents the function $g(T)=\beta T^3$, where $\beta=100$. The red line represents the function $g(T)=\beta T^2$, where $\beta= 100$.}
\label{fig5}
\end{figure}

\begin{figure*}[t]
\begin{center}
 \subfloat{\includegraphics[width=10cm]{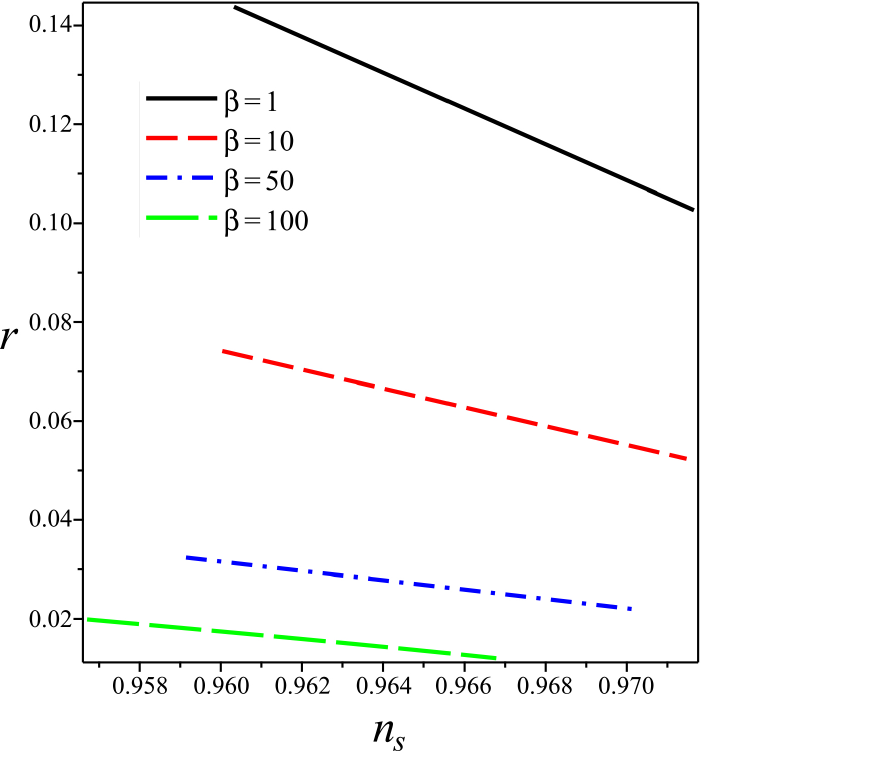}}
 \subfloat{\includegraphics[width=8.7cm]{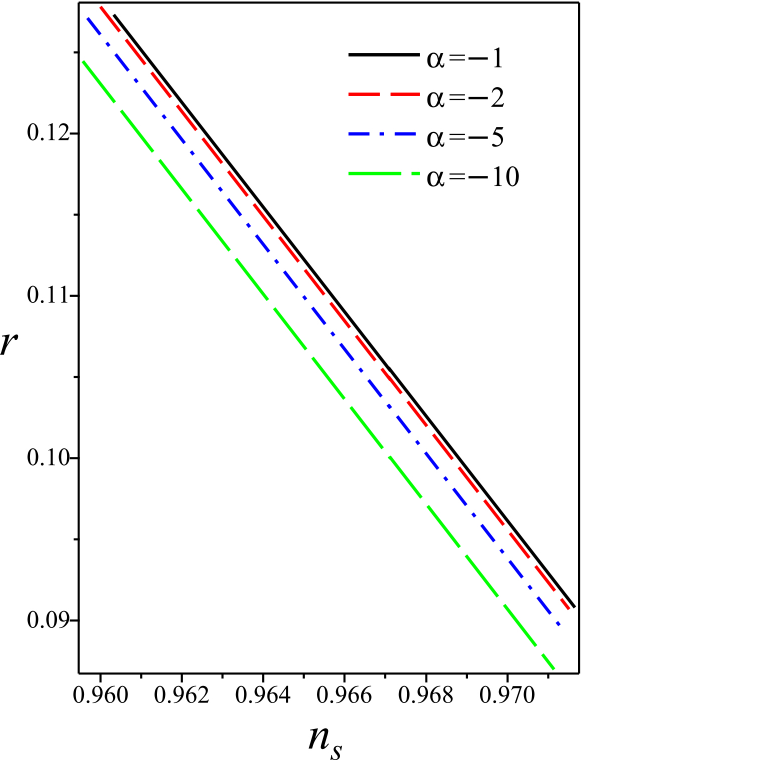}}
  \end{center}
   \caption{\footnotesize The tensor-to-scalar ratio $r$ versus the scalar spectral index $n_s$ is depicted for a quadratic potential $V(\phi)=M\phi^2$ and the function $g(T)=\beta T^2$. The left panel shows the case with $\alpha=-1$, $M=10^{-6}$, and different values of $\beta$, while the right panel shows the case with $\beta=2$, and $M=10^{-6}$ and different values of $\alpha$.Additionally, for two of the plots, the number of e-folds falls within the range of $50 < N < 70$.}
\label{fig6}
\end{figure*}

\section{Numerical discussion}

To confirm the results from the previous section, we will now perform a numerical analysis. We will investigate the quadratic potential defined by $ V(\phi) = M \phi^2$ in the context of a power law function $g(T) = \beta T^m $. Therefore we can identify four essential parameters in this model: $\alpha$ in the function $f(Q,T) = \alpha Q + g(T)$, $M$ in the quadratic potential $V(\phi) = M \phi^2$, and $m$ and $\beta$ in the power law function $g(T)$, represented as $g(T) = \beta T^m$.
Furthermore, there are six quantities associated with the universe's background during inflation: the slow roll parameters $\epsilon$, $\theta$, $\lambda$, $\eta$, $\delta$, and the number of e-folds $\mathcal{N}$.
Moreover, there are three quantities associated with linear perturbations: the scalar power spectrum $\mathcal{P}_s$, the spectral index $n_s$, and the tensor-to-scalar ratio $r$, which we can compare with observational data.
We derive the background and perturbed quantity in a general form of the function $g(T)$. Now, let's check out some specific versions of $g(T)$: $\beta T$, $\alpha T^2$, $\beta T^3$, and $\beta T^4$.

In the specific scenario where $g(T) = \beta T$ as outlined in relation \eqref{93}, we have noted that the spectral index $n_s$ and the tensor-to-scalar ratio $r$ match those observed in general relativity. This indicates that the observational constraints in this model do not depend on the parameters $\alpha$ and $\beta$. This case was examined numerically in \cite{Shiravand2022}, so we will refrain from revisiting it here.
Consequently, we will conduct a detailed analysis of the specific form $ f(Q,T)=\alpha Q+\beta T^2$ and juxtapose these findings with the observational data.
The evolution of the e-folding number $\mathcal{N}$ as a function of the scalar field $\phi$ is illustrated in FIG. \ref{fig1} for three different forms of the function $g(T)$: $\beta T$, $\beta T^2$, and $\beta T^3$. In this plot, we set $\alpha = -1$ and $M = 10^{-6}M_p$. The left panel corresponds to $\beta = 2$, while the right panel corresponds to $\beta = 200$. We have noted that the evolution of the e-folding number in the linear model $g(T)=\beta T$ occurs more rapidly than in other cases. Consequently, the end of inflation takes place at higher values of the scalar field in nonlinear models that include terms like $\beta T^2$ and $\beta T^3$.     

In figure FIG. \ref{fig2}, we plotted the e-folding $\mathcal{N}$ as a function of the scalar field $\phi$ for the quadratic potential $V(\phi)=M \phi^2$ and a function defined as $g(T)=\beta T^2$ to highlight the influence of the parameters $\alpha$, $\beta$ and $M$. In the left panel of Figure \ref{fig2}, we show the effects of different values of the parameter $\alpha$. The middle panel illustrates the impact of varying values of the parameter $\beta$. Lastly, the right panel presents the scenario with several different values of the parameter $M$.

In FIG. \ref{fig3}, we present three plots illustrating the relationship between the slow roll parameter $\epsilon$ and the scalar field $\phi$, highlighting the influence of the parameters $\alpha$, $\beta$, and $M$.
It is evident that an increase in the value of $\alpha$ results in a reduction of the slope in the $\epsilon$ versus $\phi$ graph, while higher values of $\beta$ and $M$ contribute to a steeper slope in the graphs.

In FIG. \ref{fig4}, we present the well-known $r-n_s$ plot for various functions of $g(T)$: $\beta T$, $\beta T^2$, $\beta T^3$, and $\beta T^4$.
These plots correspond to a quadratic potential given by $V(\phi)=M\phi^2$, where we have chosen $\alpha=-1$ and $M=10^{-6} M_p$.
The left panel is plotted for $\beta = 2$, and the right panel is plotted for $\beta=100$. Additionally, all curves in both panels correspond to an e-folding number within the range of $60<\mathcal{N}<70$.

These plots reveal two important insights: first, the lowest curve, represented by the red dashed line, corresponds to $\beta T^2$, while all other curves lie above this reference line. Second, it is significant to note that increases in the parameter $\beta$ result in a downward shift of the curves.
 
These plots reveal two important insights: the lowest curve (red dashed line) corresponds to $\beta T^2$, and increasing the parameter $\beta$ shifts the curves downward, with all others above this reference.

To compare the results of this model with observational data, we plotted our curves against the Planck 2018 background data in Figure FIG. \ref{fig5}.
In this figure, the solid black line represents the linear model $\beta T$. We observe that this line does not depend on the values of $\beta$, as predicted by equation \eqref{93}.
Furthermore, the solid red curve represents $g(T)=\beta T^2$ with $\beta = 100$, situated within the area defined by $\text{Planck TT, TE, EE} + \text{lowE} + \text{lensing} + \text{BAO} + \text{BICEP2/Keck}$. 
We can indicate that the model represented by $g(T)=\beta T^2$ aligns remarkably well with the observational data.

Consequently, we analyze the impact of the parameters $\alpha$ and $\beta$ on the $r-n_s$ plot for the function $g(T)=\beta T^2$, as illustrated in Figure \ref{fig6}.
We have noted that changes in the parameter $\alpha$ significantly affect the $r-n_s$ plots, whereas increases in the parameter $\beta$  lead to a better fit with the observations.
At the CMB scale $k_\star=0.05Mpc^{-1}$ the results of Planck 2018 \cite{Aghanim2021} provide the following constraints on the power spectrum, the scalar spectral index and the tensor to scalar ratio as
\begin{eqnarray}
\label{101}
\ln(10^{10}\mathcal{P}_\mathcal{R})&=&3.044\pm0.014 \;\;\;\; (68\%\; C.L.),\\
\label{102}
n_s&=&0.9649 \pm 0.0042\;\;\; (68\% \; C.L.), \\
\label{103}
r&<&0.07\;\;\;\;\;\; (95\%\; C.L.).
\end{eqnarray}

By evaluating the parameter $M$, $\alpha$ and $\beta$, we can determine the value of the scalar field at the end of inflation by applying the condition  $\epsilon(\phi_{\text{end}})=1$ in equation \eqref{34}.
To achieve appropriate e-folding number $\mathcal{N}$, we determine $\phi_\star$ by utilizing the relation presented in \eqref{37}. 
Consequently, the spectral index and the tensor-to-scalar ratio can be derived from equations \eqref{79} and \eqref{86}.
\begin{center}
\begin{tabular}[t]{c c c c c c c c c|c|}
\multicolumn{9}{ c }{\footnotesize Table 1:Model's parameters in $f(Q,T)=\alpha Q+\beta T^m$} \\
\hline
~$m$~ &~$\alpha$~ &~~$\beta$~~& ~~$\phi_{end}$~~&~~$\phi_{\star}$~~&~~$\mathcal{N}$~~ &~$n_s$~&~$r$~ \\
\hline
$1$ & $-1$ & $0$ & $1.414$&$15.556$& $60$ & $0.967$&$0.131$\\
\hline
$1$ & $-1$ & $2$  & $0.816$&$8.980$& $60$ & $0.967$&$0.132$ \\
\hline
$1$ & $-1$ & $10$  & $0.426$&$4.690$& $60$ & $0.967$&$0.132$ \\
\hline
$2$ & $-1$ & $2$  & $1.414$&$15.556$& $60$ & $0.967$&$0.106$ \\
\hline
$2$ & $-1$ & $10$  & $1.414$&$15.555$& $60$ & $0.967$&$0.061$ \\
\hline
$2$ & $-1$ & $100$  & $1.413$&$15.425$& $60$ & $0.964$&$0.014$ \\
\hline
$3$ & $-1$ & $2$  & $1.414$ &$15.556$& $60$ & $0.967$& $0.107$ \\
\hline
$3$ & $-1$ & $100$  & $1.414$&$15.557$& $60$ & $0.967$& $0.022$ \\
\hline
\end{tabular}
\end{center}

It is evident that an appropriate choice of the model's parameters enables $f(Q,T)$ gravity, with $g(T)=\beta T^2$, to align closely with the observational data from Planck 2018 \cite{Aghanim2021}.
The energy density at the time of horizon crossing can be approximated by the expression $V(\phi_\star)=M \phi_\star^2$. For the specific case where $\beta T^2$ with $\alpha=-1$ and $\beta=2$, this results in $\rho_\star \approx 2.25 \times 10^{-4}$. This value is an order of magnitude below the Planck energy scale, which is consistent with the energy scale relevant to big bang cosmology.
\section{Conclusion}

In this paper, we have examined inflation and the generation of primordial fluctuations within the framework of $f(Q,T)$ gravity , specified by the functional form $f(Q,T)=\alpha Q+g(T)$, where $\alpha$ is a constant value and $g(T)$ is an arbitrary function of the trace of the energy-momentum tensor $T$. We explore slow roll inflation within this framework by deriving the slow-roll parameters and e-folding number as functions of the inflaton potential for a general form of the function $g(T)$.
Furthermore, to investigate the generation of primordial fluctuations, we derive the Mukhanov-Sasaki equations for both scalar and tensor modes in the uniform gauge. By solving these equations and applying asymptotic solutions, we can determine the scalar power spectrum, the scalar spectral index, and the tensor-to-scalar ratio in this framework. 
Additionally, this model does not encounter any issues with the graceful exit problem, in contrast to many other models that require manual integration of graceful exit functionality.
For instance, we explored spatial scenarios in which the function $g(T)$ is represented as $\beta T$ and $\beta T^2$. In these cases, we calculated various important observational quantities, such as the spectral index and the tensor-to-scalar ratio, specifically for the quadratic potential $V(\phi)=M\phi^2$.
In the case where $g(T)=\beta T$, we have noted that the results are the same as those found in general relativity. The generation of observable quantities in the case $\beta T^2$ aligns with the $2018$ Planck data appropriately.

Furthermore, to analyze the evolution of the inflationary quantity in this model, we generated plots for several parameters, such as the number of e-folds $\mathcal{N}$, the slow roll parameter $\epsilon$, the scalar spectral index $n_s$, and the tensor-to-scalar ratio $r$ with the different values of our model parameters $\alpha$, $\beta$, $M$, and $m$.
We found that in the $\beta T^2$ scenario the slow roll parameter $\epsilon$, is sufficiently small of order $\sim\mathcal{O}(0.01)$ and maintains an appropriate value during the entire inflationary phase.
 Additionally, the total number of e-folds in this model, from the beginning to the end of inflation, can be sufficient to address the issues of the standard model, including the horizon problem and others. 
Also, the graphs depicting the tensor-to-scalar ratio in relation to the scalar spectral index indicate that these parameters align with the observational data of Planck 2018 when considering $g(T) = \beta T^2$. 
Furthermore, to assess the performance of this model in relation to observational data, we generated plots of our curves for $(r,n_s)$ utilizing the Planck 2018 background data for $\beta T$, $\beta T^2$, and $\beta T^3$. Our analysis revealed that these curves are appropriately situated within the area delineated by $\text{Planck TT, TE, EE} + \text{lowE} + \text{lensing} + \text{BAO} + \text{BICEP2/Keck}$.
 
We can assert that the model described by $g(T)=\beta T^2$ fits the observational data exceptionally well and serves as an effective framework for comprehending the background and fluctuations of inflation..

\end{document}